\documentclass[final,5p,times,twocolumn]{elsarticle}

\usepackage{graphicx}

\usepackage{amssymb}

\usepackage{epsfig}

\begin{document}

\begin{frontmatter}




\title{Progenitors of type Ia supernovae}

\author[label1,label2]{Bo Wang}
\ead{wangbo@ynao.ac.cn}
\author[label1,label2]{Zhanwen Han}
\ead{zhanwenhan@ynao.ac.cn}
\address[label1]{National Astronomical Observatories/Yunnan
Observatory, Chinese Academy of Sciences, Kunming 650011, China}
\address[label2]{Key
Laboratory for the Structure and Evolution of Celestial Objects,
Chinese Academy of Sciences,  Kunming 650011, China}

\begin{abstract}

Type Ia supernovae (SNe~Ia) play an important role in
astrophysics and are crucial for the studies of stellar evolution, galaxy
evolution and cosmology. They are generally thought to be
thermonuclear explosions of accreting carbon--oxygen white dwarfs (CO WDs) in
close binaries, however, the nature of the mass donor star is still
unclear. In this article, we review various progenitor models
proposed in the past years and summarize many observational
results that can be used to put constraints on the nature of their
progenitors. We also
discuss the origin of SN~Ia diversity and the impacts of SN~Ia
progenitors on some fields. The currently favourable progenitor
model is the single-degenerate (SD) model, in which the WD accretes material from a non-degenerate companion star.
This model may explain the similarities of most SNe Ia.
It has long been argued that the double-degenerate (DD) model,
which involves the merger of two CO WDs, may lead to an accretion-induced
collapse rather than a thermonuclear explosion.
However, recent observations of a few SNe Ia seem to support the DD model, and this model
can produce normal SN Ia explosion under certain conditions.
Additionally, the sub-luminous SNe Ia may be explained by the sub-Chandrasekhar mass model. At present, it
seems likely that more than one progenitor model, including some variants of the SD and DD models, may be required to explain the observed diversity of SNe Ia.

\end{abstract}


\end{frontmatter}


\section{Introduction}

Type Ia supernova (SN~Ia) explosions are among the most energetic
events observed in the Universe. They are defined as those without
hydrogen or helium lines in their spectra, but with strong SiII absorption
lines around the maximum light (Filippenko, 1997). They appear to be good cosmological
distance indicators due to their high luminosities and remarkable
uniformity, and thus are used for determining the cosmological
parameters (e.g. \textbf{$\Omega_{M}$} and
\textbf{$\Omega_{\Lambda}$}; Riess et al., 1998; Perlmutter et al.,
1999). This leads to the discovery of the accelerating expansion of
the Universe that is driven by the mysterious dark energy. SNe Ia are also a
key part of our understanding of galactic chemical evolution owing
to the main contribution of iron to their host galaxies (e.g.
Greggio and Renzini, 1983; Matteucci and Greggio, 1986). In addition,
they are accelerators of cosmic rays and as sources of kinetic energy
in galaxy evolution processes (e.g. Helder et al., 2009; Powell
et al., 2011). The use of SNe Ia as standard candles
is based on the assumption that all SNe Ia have similar
progenitors and are highly homogeneous. However,
several key issues related to the nature of their progenitors and explosion
mechanism are still not well understood (Branch et al., 1995; Hillebrandt and
Niemeyer, 2000). This may directly affects the reliability of the results of the current
cosmological model and galactic chemical evolution model.

When SNe~Ia are used as distance indicators, the Phillips relation
is adopted, which is a phenomenological linear relation between the
absolute magnitude of SNe~Ia and the magnitude difference from its $B$-band maximum to 15 days after that
(Phillips, 1993).
The Phillips relation is based on the SN~Ia sample of low
redshift Universe ($z<0.05$) and assumed to be valid at high
redshift. This assumption is precarious since there is still no
agreement on the nature of their progenitors. If the properties of
SNe~Ia evolve with the redshift, the results for cosmology might be
different. In addition, more observational evidence
indicates that not all SNe Ia obey the Phillips relation (e.g. Wang et al., 2006).

Aside from the Phillips relation, many updated versions of this method are given to
establish the relation between SN Ia intrinsic luminosities and the shape of their light curves.
The stretch factor $s$ method was proposed to measure the
light curve shape by adjusting the scale on the time axis by a
multiplicative factor (Perlmutter et al., 1997; Goldhaber et al.,
2001). In addition, an empirical method based on multicolor light curve
shapes has been developed to estimate the luminosity,
distance, and total line-of-sight extinction of SNe Ia (Riess et
al., 1998). Wang et al. (2005) presented a single post-maximum
color parameter $\Delta C_{12}$ ($B - V$ color $\sim$12 days after
the $B$-band light maximum), which empirically describes almost the
full range of the observed SN Ia luminosities and gives tighter
correlations with their luminosities, but the underpinning physics
is still not understood. Recently, Guy et al. (2005) used an innovative approach to constrain
the spectral energy distribution of SNe Ia, parameterized
continuously as a function of color and stretch factor $s$, and
allow for the generation of light curve templates in arbitrary
pass-bands. This method was known as the spectral adaptive light
curve template method, which offers several practical
advantages that make it easily applicable to high redshift SNe Ia. The
k-corrections are built into the model but not applied to the
data, which allows one to propagate all the uncertainties directly
from the measurement errors.

It is widely accepted that SNe~Ia arise from thermonuclear
explosions of carbon--oxygen white dwarfs (CO WDs) in
close binaries (Hoyle and Fowler, 1960; Nomoto et al., 1997). This hypothesis is supported
by the fact that the amount of energy observed in SN explosions is equal to the
amount that would be produced in the conversion of carbon and
oxygen into iron ($\sim$$10^{51}$\,erg; Thielemann et al., 2004).
The energy released from the nuclear burning completely
destroys the CO WD and produces a large amount of $^{56}$Ni.
The optical/infrared light curves are powered by
the radioactive decay of $^{56}$Ni $\rightarrow$ $^{56}$Co
$\rightarrow$ $^{56}$Fe. In order to trigger the carbon ignition,
the mass of the CO WD must grow close to the Chandrasekhar (Ch) mass.
When the WD increases its mass close to the Ch mass, it is thought to ignite
near the center; at first the flame propagates subsonically
as a deflagration, and in a second phase a detonation
triggers, which propagates supersonically and completely destroys the CO WD (Hillebrandt and
Niemeyer, 2000). The
realistically conceivable way to make the WD grow to the Ch mass is via
mass-transfer from a mass donor star in a close binary.
However, the nature of the mass donor
star in the close binary is still uncertain, and no progenitor
system before SN explosion has been conclusively identified. Additionally, there
is some observational evidence that a subset of SNe~Ia have progenitors with
a mass exceeding or below the standard Ch mass limit (e.g.
Howell et al., 2006; Foley et al., 2009; Wang et al., 2008a).

Many progenitor models of SNe~Ia have been proposed in the past years.
The most popular progenitor models are single-degenerate (SD)
and double-degenerate (DD) models.
In Sect. 2, we review various progenitor models, including some variants of the SD and DD models proposed in the
literature. We summarize some observational ways to test the current
progenitor models in Sect. 3, and introduce some objects that may be related to the
progenitors and the surviving companion stars of SNe Ia in Sect. 4.
We discuss the origin of SN~Ia diversity and the impacts of SN~Ia
progenitors on some research fields in Sects. 5 and 6, respectively. Finally, a summary is given in Sect.
7. For more discussions on these subjects, see previous reviews on SN Ia progenitors (e.g. Branch et al., 1995; Hillebrandt and
Niemeyer, 2000; Livio, 2000; Nomoto et al., 2003; Podsiadlowski, 2010; Maoz and Mannucci, 2012).

\section{Progenitor models}
\subsection{Single-degenerate model}
In this model, a CO WD accretes hydrogen-rich or helium-rich
material from a non-degenerate companion star, increases its mass to
the Ch mass, and then explodes as a SN~Ia (Whelan and Iben, 1973;
Nomoto, 1982a). The SD model may explain the similarities of most SNe Ia, since
SN Ia explosions in this model occur when the CO WD increases its
mass to the maximum stable mass (i.e. the Ch mass). In addition,
the observed light curves and early time spectra of the majority of SNe Ia are in
excellent agreement with the synthetic spectra of the SD Ch mass model
(Nomoto et al., 1984; H\"{o}flich et al., 1996; Nugent et al., 1997).

The companion star in this model could be a main-sequence (MS) star
or a subgiant star (WD + MS channel), or a red-giant star
(WD + RG channel), or a helium star (WD + He star channel) (Hachisu et al., 1996,
1999a,b; Li and van den Heuvel, 1997; Langer et al., 2000; Han and Podsiadlowski, 2004; Fedorova et al., 2004;
Meng et al, 2009; Wang et al., 2009a, 2010a). The main
problem for this class of models is that it is generally difficult
to increase the mass of the WD by accretion. Whether the WD can grow
in mass depends crucially on the mass-transfer rate and the
evolution of the mass-transfer rate with time. (1) If the rate is too
high, the system may enter into a common envelope (CE) phase; (2) if
the rate is too low, the nuclear burning is unstable that leads to nova explosions
in which all the accreted matter is ejected. There is only a very
narrow parameter range in which the WD can accrete H-rich or
He-rich material and burn in a stable manner. This
parameter range may be increased if the rotation
affects the WD mass-accretion process (Yoon and Langer, 2004).

An essential element in this
model is the optically thick wind assumption, which
enlarges the parameter space for producing SNe Ia (Hachisu et al., 1996,
1999a,b; Li and van den Heuvel, 1997; Han and Podsiadlowski, 2004;
Wang et al., 2009a, 2010a).
In this assumption, taking a MS donor star for an example, if the mass-transfer rate from the MS star
exceeds a critical value, $\dot{M}_{\rm cr}$, it is assumed that the accreted H
burns steadily on the surface of the WD and that the H-rich
material is converted into He at the rate of $\dot{M}_{\rm cr}$.
The unprocessed matter is
assumed to be lost from the binary system as an optically thick wind. However, this assumption is very
sensitive to the Fe abundance, and it is likely
that the wind does not work when the metallicity is lower than a certain
value.\footnote{At low enough metallicities (e.g. $Z<0.002$), the optical depth of the wind would become
small, and thus the wind-regulation mechanism
would become ineffective (e.g. Kobayashi et al., 1998; Kobayashi and Nomoto, 2009).
In this case, the binary system will pass through a CE phase before
reaching the Ch mass. Thus, if this is true then there would be an
obvious low-metallicity threshold for SNe Ia in comparison with SNe
II. However, the metallicity threshold has not been found in
observations (Prieto et al., 2008; Badenes et al., 2009a).}

\subsubsection{WD + MS channel}
\begin{figure*}
\centerline{\epsfig{file=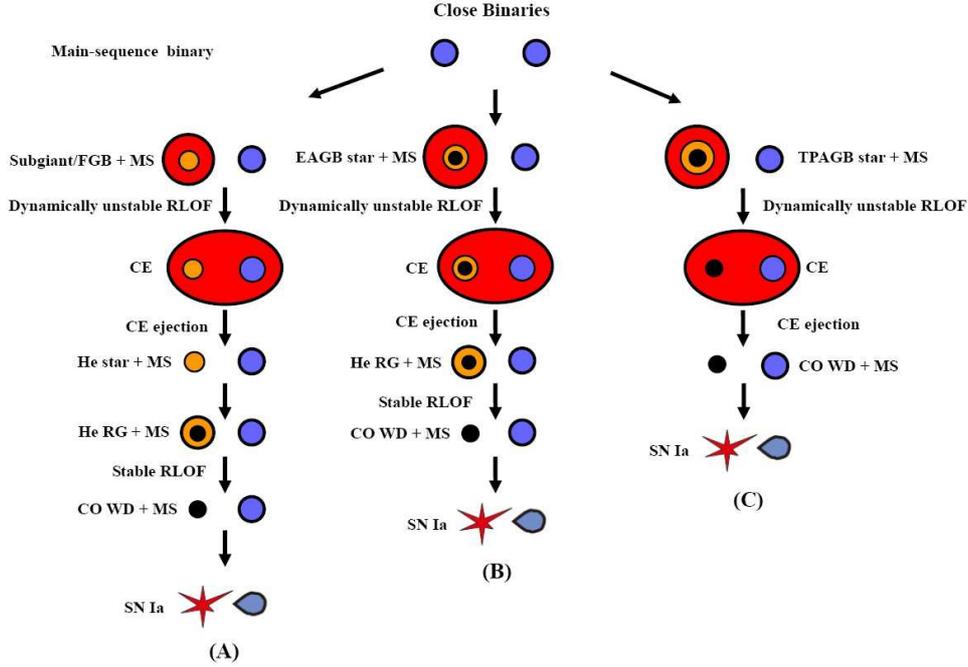,angle=0,width=13.0cm}}
\caption{Binary evolutionary scenarios of the WD + MS channel for producing SNe Ia.}
\end{figure*}

In the WD + MS channel (usually called the supersoft channel), a CO
WD in a binary system accretes H-rich material from a MS or a
slightly evolved subgiant star. The accreted H-rich material
is burned into He, and then the He is converted to carbon and
oxygen. When the
CO WD increases its mass close to the Ch mass, it explodes as a SN~Ia.
Based on the evolutionary phase of the primordial primary (i.e. the massive star) at the
beginning of the first Roche lobe overflow (RLOF), there are three
evolutionary scenarios to form WD + MS systems and then produce
SNe~Ia (Fig. 1; for details see Wang et al., 2010a;
also see Postnov and Yungelson, 2006; Meng et al, 2009).

\textbf{Scenario A:} The primordial
primary first fills its Roche lobe when it is in the Hertzsprung gap
(HG) or first giant branch (FGB) stage (i.e. Case B mass-transfer defined by
Kippenhahn and Weigert, 1967). In this case, due to a large
mass-ratio or a convective envelope of the mass donor star, a CE may be formed (Paczy\'{n}ski, 1976). After
the CE ejection, the primary becomes a He star and continues to
evolve. After the exhaustion of central He, the He star evolves to the
RG stage. The He RG star that now
contains a CO-core may fill its Roche lobe again due to the
expansion itself, and transfer its remaining
He-rich envelope onto the surface of the MS companion star,
eventually leading to the formation of a CO WD + MS system.
For this scenario, SN Ia explosions occur for the
ranges $M_{\rm 1,i}\sim4.0$$-$$7.0\,M_\odot$, $M_{\rm
2,i}\sim1.0$$-$$2.0\,M_\odot$, and $P^{\rm i} \sim 5$$-$30\,days,
where $M_{\rm 1,i}$, $M_{\rm 2,i}$ and $P^{\rm i}$ are the initial
masses of the primary and the secondary at zero age main-sequence (ZAMS), and the initial
orbital period of the binary system.

\textbf{Scenario B:}  If the primordial primary is on the early
asymptotic giant branch (EAGB, i.e. He is exhausted in the center of
the star while this star has a thick He-burning layer and the
thermal pulses have not yet started), a CE may
be formed due to the dynamically unstable mass-transfer. After the
CE is ejected, a close He RG + MS binary may be produced;
the binary orbit decays in the process of the CE ejection and the primordial primary may evolve to a
He RG that contains a CO-core.
The He RG may fill its Roche lobe and start mass-transfer,
which is likely stable and results in a CO WD + MS system. For this scenario,
SN Ia explosions occur for the ranges $M_{\rm
1,i}\sim2.5-6.5\,M_\odot$, $M_{\rm 2,i}\sim1.5-3.0\,M_\odot$ and
$P^{\rm i} \sim 200-900$\,days.

\textbf{Scenario C:}  The primordial primary fills its Roche lobe at
the thermal pulsing asymptotic giant branch (TPAGB) stage. A CE is
easily formed owing to the dynamically unstable mass-transfer during
the RLOF. After the CE ejection, the primordial primary becomes a CO
WD, then a CO WD + MS system is produced. For this scenario, SN Ia
explosions occur for the ranges $M_{\rm
1,i}\sim4.5$$-$$6.5\,M_\odot$, $M_{\rm
2,i}\sim1.5$$-$$3.5\,M_\odot$, and $P^{\rm i}>1000$\,days.

Among the three evolutionary scenarios above, models predict that
scenario A is the more significant route for producing
SNe Ia (e.g. Wang et al., 2010a).
The WD + MS channel has been identified in recent years as supersoft
X-ray sources and recurrent novae (van den Heuvel et al., 1992;
Rappaport et al., 1994; Meng and Yang,
2010a). Many works have been concentrated on this channel. Some authors
studied the WD + MS channel with a simple analytical method to treat
binary interactions (e.g. Hachisu et al., 1996, 1999a, 2008).
Such analytic prescriptions may not describe some mass-transfer phases well enough,
especially those occurring on a thermal time-scale.
Li and van den
Heuvel (1997) studied this channel from detailed binary evolution
calculation with two WD masses (e.g. 1.0 and 1.2\,$M_{\odot}$).
Langer et al. (2000) investigated this channel for
metallicities $Z=0.001$ and 0.02, but they only studied Case A
evolution (mass-transfer during the central H-burning phase). Han
and Podsiadlowski (2004) carried out a detailed study of this
channel including Case A and early Case B for $Z=0.02$. The Galactic SN Ia birthrate from this study is
$0.6-1.1\times10^{-3}\,{\rm yr}^{-1}$. Following
the studies of Han and Podsiadlowski (2004), Meng et al. (2009)
studied the WD + MS channel comprehensively and systematically at
various metallicities.

King et al. (2003) inferred that the mass-accretion rate on to the WD
during dwarf nova outbursts can be sufficiently high to allow steady
nuclear burning of the accreted matter and growth of the WD mass.
Recently, Xu and Li (2009) also emphasized that, during the mass-transfer
through the RLOF in the evolution of WD binaries, the accreted material
can form an accretion disc surrounding the WD, which may become
thermally unstable (at least during part of the mass-transfer
lifetime), i.e. the mass-transfer
rate is not equivalent to the mass-accretion rate onto the WD.
By
considering the effect of the thermal-viscous instability of
accretion disk on the evolution of WD binaries, Wang et al. (2010a)
recently enlarged the regions of the WD + MS channel for producing
SNe~Ia, and confirmed that WDs in this channel with an initial mass as low as
$0.6\,M_\odot$ can accrete efficiently and reach the Ch mass limit.
Based on a detailed binary population synthesis (BPS)
approach,\footnote{BPS is a useful tool to simulate a large population of
stars or binaries and can help understand processes that are
difficult to observe directly or to model in detail (e.g. Han et al., 1995;
Yungelson and Livio, 2000; Nelemans et al., 2001).} they found that this channel is effective for producing
SNe~Ia (up to $1.8\times10^{-3}\,{\rm yr}^{-1}$ in the Galaxy),
which can account for about $2/3$ of the observations (see also Meng
and Yang, 2010a). However,
the parameter regions for producing SNe Ia in this model depend on
many uncertain input parameters, in particular the duty cycle
during the nova outbursts that is still poorly known.
Additionally, whether dwarf nova outbursts can increase the mass of
a WD is still a problem (e.g. Hachisu et al., 2010).

\subsubsection{WD + RG channel}
\begin{figure}[tb]
\begin{center}
\includegraphics[width=6.0cm,angle=0]{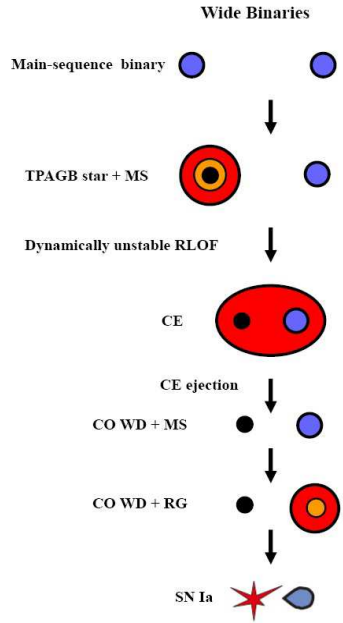}
 \caption{Similar to Fig. 1, but for the WD + RG channel.}
  \end{center}
\end{figure}

The mass donor star in this channel is a RG star, which is also called
the symbiotic channel. There is one evolutionary scenario that can
form WD + RG binaries and then produce SNe~Ia (Fig. 2;
for details see Wang et al, 2010a). Compared with the WD + MS channel,
SNe Ia in the WD + RG channel are from wider primordial binaries.
The primordial primary fills its Roche lobe at the TPAGB stage. A CE is
easily formed due to the dynamically unstable mass-transfer during
the RLOF. After the CE ejection, the primordial primary becomes a CO
WD. The MS companion star
continues to evolve until the RG stage, i.e. a CO WD + RG binary is
formed. For the WD + RG systems, SN Ia explosions
occur for the ranges $M_{\rm 1,i}\sim5.0$$-$$6.5\,M_\odot$, $M_{\rm
2,i}\sim1.0$$-$$1.5\,M_\odot$, and $P^{\rm i}>1500$\,days.

Unfortunately, the WD + RG binary usually undergoes a CE
phase when the RG star overflows its Roche lobe. More importantly, the
appropriate initial parameter space for producing SNe Ia in this channel is too
small. Thus, WD + RG
binaries seem to unlikely become a major way to form SNe~Ia. Many
authors claimed that the SN~Ia birthrate via the WD + RG channel is
much lower than that from the WD + MS channel (Yungelson and Livio,
1998; Han and Podsiadlowski, 2004; L\"{u} et al., 2006; Wang et al.,
2010a). The lowest initial WD mass in this channel for producing SNe Ia
is about $1.0\,M_\odot$ (e.g. Wang and Han, 2010a). In order to stabilize the mass-transfer process and avoid
the formation of the CE, Hachisu et al. (1999b) assumed a
mass-stripping model in which a stellar wind from the WD collides
with the RG surface and strips some of the mass from
the RG.
They obtained a high SN Ia birthrate ($\sim$$0.002\,{\rm
yr}^{-1}$) for this channel.
Here, Hachisu et al. (1999b) used
equation (1) of Iben and Tutukov (1984) to estimate the birthrate, i.e.
\begin{equation}
\nu = 0.2\,\Delta q \int_{M_{\rm A}}^{M_{\rm B}} {{d M} \over
M^{2.5}} \Delta \log A \, \mbox{yr}^{-1},
\label{realization_frequence}
\end{equation}
where $\Delta q$, $\Delta \log A$, $M_{\rm A}$ and $M_{\rm B}$ are
the appropriate ranges of the initial mass ratio, the initial
separation, and the lower and upper limits of the primary mass for
producing SNe Ia in units of solar masses, respectively. However, the birthrate is
probably overestimated, since some parameter spaces considered to
produce SNe~Ia in equation (1) may not contribute to SNe~Ia.

In symbiotic systems, WDs can accrete a fraction of the stellar wind from
cool giants. It is generally believed that the stellar wind from a normal RG is expected to be
largely spherical owing to the spherical stellar surface and isotropic
radiation. However, the majority ($>$$80\%$) of the observed planetary
nebulae are found to have aspherical morphologies (Zuckerman
and Aller, 1986). Additionally, the stellar winds from cool giants in symbiotic systems
flow out in two ways: an equatorial disc and a spherical wind.
In this context, by assuming an aspherical stellar wind with an equatorial disk from
a RG, L\"{u} et al. (2009) investigated the production of
SNe~Ia via the symbiotic channel. They estimated that the Galactic
SN~Ia birthrate via this channel is between $2.27\times
10^{-5}$\,yr$^{-1}$ and $1.03\times 10^{-3}$, and the theoretical
SN~Ia delay time (between the star formation and SN explosion) has a
wide range from 0.07 to 5\,Gyr. However, these results are greatly
affected by the outflow velocity and the mass-loss rate of the
equatorial disk.

The stellar wind from RG stars might be enhanced by tidal or other
interactions with a companion. Tout and Eggleton (1988) brought
the tidally enhanced stellar wind assumption to explain the mass inversion in RS CVn
binaries. This assumption has been widely used to explain
many phenomena related to giant star evolution in binaries (e.g. Han, 1998; van Winckel, 2003).
The tidally enhanced stellar wind assumption has two advantages in the studies of symbiotic systems: (1) The
WD may grow in mass substantially by accretion from stellar
wind before RLOF; (2) the mass-transfer may be stabilized
because the mass ratio ($M_{\rm giant}$/$M_{\rm WD}$) can be much reduced at the
onset of RLOF. By adopting the
tidally enhanced stellar wind assumption, Chen et al. (2011) recently argued that the parameter space of SN~Ia progenitors
can be extended  to longer
orbital periods for the WD + RG channel (compared to the
mass-stripping model of Hachisu et al., 1999b), and thus increase
the birthrate up to $6.9\times 10^{-3}$\,yr$^{-1}$, which is
also probably overestimated due to the use of equation (1).
Additionally, the parameter space of SN~Ia
progenitors strongly depends on the tidal wind enhancement parameter
$B_{\rm w}$ that is still poorly known.

In a variant of the symbiotic channel, the mass-transfer from carbon-rich AGB stars with WD
components can occur via stellar winds or RLOF (Iben and Tutukov,
1985). It has been suggested that an AGB donor star is in the progenitor system of SN 2002ic,
which is an atypical SN~Ia with evidence for substantial amounts of
hydrogen associated with the system (Hamuy et al., 2003).
Recently, Chiotellis et al. (2012) presented a WD with an
AGB donor star for the SN remnant (SNR) of SN 1604, also known as
Kepler's SNR. They argued that its main features can be explained by
the model of a symbiotic binary consisting of a WD and an AGB donor
star with an initial mass of 4$-$5\,$M_{\odot}$. Detailed
calculations of binary evolutionary model are needed to understand
whether these WD components in  WD + AGB binaries can result in
SN~Ia explosions.

\subsubsection{WD + He star channel}

\begin{figure*}
\centerline{\epsfig{file=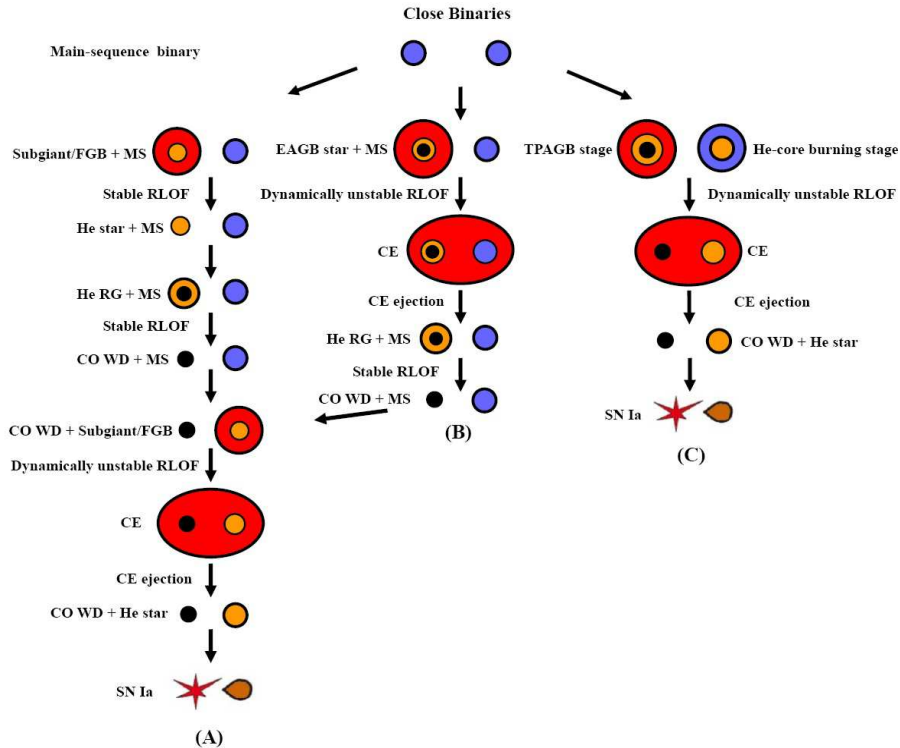,angle=0,width=13.0cm}}
\caption{Similar to Fig. 1, but for the WD + He channel.}
\end{figure*}

A CO WD may also accrete helium-rich material from a He star or
a He subgiant to increase its mass to the Ch mass, which is also
known as the He star donor channel. There are three evolutionary
scenarios to form WD + He star systems and then produce SNe~Ia (see
Fig. 3; for details see Wang et al., 2009b).

\textbf{Scenario A:}
The primordial primary first fills its Roche lobe when it is in the
HG or FGB stage. At the end of the RLOF, the primary becomes a He
star and continues to evolve. After the exhaustion of central He,
the He star evolves to the RG stage. The He RG star that now contains a CO-core may fill its Roche lobe
again due to the expansion of itself, and transfer its
remaining He-rich envelope to the MS companion star, eventually
leading to the formation of a CO WD + MS system. After that, the MS
companion star continues to evolve and fills its Roche lobe in the
HG or FGB stage. A CE is possibly formed due to the
dynamically unstable mass-transfer. If the CE can be ejected, a
close CO WD + He star system is then produced. The CO WD + He star
system continues to evolve, and the He star may fill its Roche lobe
again (due to the orbit decay induced by the gravitational wave
radiation or the expansion of the He star itself), and transfer some
material onto the surface of the CO WD. The accreted He may be
converted into carbon and
oxygen via the He-shell burning, and the CO WD
increases in mass and explodes as a SN Ia when its mass reaches the
Ch mass. For this scenario, SN Ia
explosions occur for the ranges $M_{\rm 1,i}\sim5.0-8.0\,M_\odot$,
$M_{\rm 2,i}\sim2.0-6.5\,M_\odot$ and $P^{\rm i} \sim 10-40$\,days.

\textbf{Scenario B:} If the primordial primary is on the
EAGB stage at the onset of the RLOF, a CE may be formed due to the dynamically unstable
mass-transfer. After the CE is ejected, a close He RG + MS binary may be produced;
the binary orbit decays in the procedure of the CE ejection and the primordial primary becomes a He RG.
The He RG may fill its Roche
lobe and start the mass-transfer, which is likely stable and results in
a CO WD + MS system. The subsequent evolution of this system is
similar to scenario A above, and may form a CO WD + He star system
and finally produce a SN Ia. For this scenario, SN Ia explosions occur for the
ranges $M_{\rm 1,i}\sim6.0-6.5\,M_\odot$, $M_{\rm
2,i}\sim5.5-6.0\,M_\odot$ and $P^{\rm i} \sim 300-1000$\,days.

\textbf{Scenario C:} The primordial
primary fills its Roche lobe at the TPAGB stage, and the companion
star evolves to the He-core burning stage. A double-core CE may be formed
owing to the dynamically unstable mass-transfer during the RLOF.
After the CE ejection, the primordial primary becomes a CO WD, and the companion
star is a He star at the He-core burning stage, i.e.
a CO WD + He star system is formed. The subsequent evolution of this system is similar to that in the above two scenarios,
i.e. a SN Ia may be produced. For this scenario, SN Ia explosions occur
for the ranges $M_{\rm 1,i}\sim5.5-6.5\,M_\odot$, $M_{\rm
2,i}\sim5.0-6.0\,M_\odot$ and $P^{\rm i}>1000$\,days.

SNe~Ia from the He star donor channel can neatly avoid H lines,
consistent with the defining spectral characteristic of most SNe~Ia.
Yoon and Langer (2003) followed the evolution of a WD + He star
binary with a $1.0\,M_{\odot}$ WD and a $1.6\,M_{\odot}$ He star in
a 0.124\,d orbit. In this binary, the WD accretes He from the He
star and grows in mass to the Ch mass. Based on the optically thick
wind assumption, Wang et al. (2009a) systematically  studied  the He star donor
channel. In the study, they carried out binary evolution
calculations of this channel for about 2600 close WD + He star
binaries. The study showed the initial parameter spaces for the progenitors
of SNe~Ia, and found that the
minimum mass of CO WD  for producing SNe Ia  in this channel may be
as low as $0.865\,M_{\odot}$. By using a detailed BPS approach, Wang et al. (2009b)
found that the Galactic SN~Ia birthrate from this channel is
$\sim$$0.3\times 10^{-3}\,{\rm yr}^{-1}$ and  this channel can
produce SNe~Ia with short delay times ($\sim$45$-$140\,Myr). Wang and Han
(2010b) also studied the He star donor channel with different
metallicities. For a constant star-formation galaxy (like our own
galaxy), they found that SN~Ia birthrates increase with metallicity.
If a single starburst is assumed (like in an elliptical galaxy), SNe~Ia
occur systematically earlier and the peak value of the birthrate is
larger for a higher metallicity.

\subsection{Double-degenerate model}
In the DD model, SNe~Ia arise from the merging of two close CO WDs
that have a combined mass larger than or equal to the Ch mass
(Tutukov and Yungelson, 1981; Iben and Tutukov, 1984; Webbink,
1984). Both CO WDs are brought together by gravitational wave radiation
on a timescale $t_{\rm GW}$ (Landau and Lifshitz, 1971),
\begin{equation}
t_{\rm GW}(\rm yr \it)=\rm 8\times10^{7}\it(\rm yr \it
)\times\frac{(M_{\rm 1}+M_{\rm 2})^{\rm 1/3}}{M_{\rm 1}M_{\rm
2}}P^{\rm 8/3}(\rm h),
\end{equation}
where $P$ is the orbital period in hours, $t_{\rm GW}$ in years and
$M_{\rm 1}$, $M_{\rm 2}$ in $M_{\odot}$.
The delay time from the star formation to the occurrence of a SN~Ia
is equal to the sum of the timescale that the secondary star
becomes a WD and the orbital decay time $t_{\rm GW}$. For the DD model, there are three binary evolutionary
scenarios to form double CO WD systems and then produce SNe~Ia, i.e. stable RLOF plus
CE ejection scenario, CE ejection plus CE ejection scenario and exposed core plus
CE ejection scenario (for details see Han, 1998).

The DD model has the advantage that the theoretically predicted
merger rate is quite high, consistent with the observed SN~Ia
birthrate (e.g. Yungelson et al., 1994; Han, 1998; Nelemans et al.,
2001; Ruiter et al., 2009; Wang et al., 2010b).\footnote{Badenes and Maoz (2012) recently calculated the merger rate of binary WDs in the Galactic disk based on
the observational data in the Sloan Digital Sky Survey. They claimed that there are not enough double WD systems
with the super-Ch masses to reproduce the observed SN Ia birthrate in the context of the DD model.}
Importantly, this
model can naturally explain the lack of H or He emission in the spectra of SNe Ia.
As an additional argument in favor of the DD model, one may consider
this model to explain some observed super-luminous SNe Ia (for more discussions see Sect. 2.3.2).
Furthermore, there are some double WD progenitor candidates that have been found in observations, and
recent observations of a few SNe Ia seem to support the DD model (for more discussions see Sect. 3).
However, the DD model has difficulties in explaining the similarities of most SNe Ia, since
the merger mass in this model varies for different
binaries and has a relatively wide range ($\sim$$1.4-2.0\,M_{\odot}$; Wang et al., 2010b).

Most importantly, the
merger of two WDs may result in an accretion-induced collapse to form a neutron
star rather than a thermonuclear explosion (Nomoto and Iben,
1985; Saio and Nomoto, 1985; Timmes et al., 1994).
In the process of a double-WD merger, once the less massive WD
fills its Roche lobe, it is likely to be disrupted and rapidly accreted
by the more massive one. Meanwhile, the less massive WD is transformed into a
disk-like structure around the more massive companion. It is usually
assumed that in this configuration the temperature maximum is located at
the ``disk-dwarf'' interface and that carbon burning starts there.
In this process, the carbon burning front propagates inward and then the CO WD is transformed
into an O-Ne-Mg WD, which collapses to form a neutron star by electron capture on $^{24}$Mg.

There may be some parameter ranges where the accretion-induced collapse can be
avoided (e.g. Piersanti et al., 2003; Yoon et al., 2007). Piersanti et al. (2003)
suggested that the double WD merger process could be quite violent, and might
lead to a SN Ia explosion under the right conditions. Pakmor et al. (2010) argued that the violent mergers of
two equal-mass CO WDs  ($\sim$$0.9\,M_{\odot}$, critical conditions for the successful initiation
of a detonation) can be obtained, and may explain the formation of
sub-luminous 1991bg-like objects. Although the light curve from the merger
model is broader than that of SN 1991bg-like objects, the synthesized
spectra, red color and low expansion velocities are all close to
those observed for SN 1991bg-like objects (Pakmor et al., 2010). In a further study,
Pakmor et al. (2011) claimed that a high mass-ratio is required
for this model to work; for a primary
mass of $0.9\,M_{\odot}$ a mass-ratio of at least about 0.8.
This result will affect the potential SN Ia birthrate of the DD model. We note that
van Kerkwijk et al. (2010) came to a similar conclusion before Pakmor et al. (2011),
but that was in turn partially based on Pakmor et al. (2010) and Lor\'{\i}n-Aguilar et al. (2009).
Adopting the results of Pakmor et al. (2011) with a detailed BPS approach, Meng et al. (2011) estimated that the sub-luminous
events from this model may only account for not more than $1\%$ of
all SNe~Ia.

Recently, by assuming that the moment at which the detonation forms is an artificial
parameter, Pakmor et al. (2012) presented a fully three-dimensional simulation of a violent merger of two CO WDs
with masses of $0.9\,M_{\odot}$ and $1.1\,M_{\odot}$, by combining very high resolution and the exact initial conditions.
They estimated that the simulation produces about $0.62\,M_{\odot}$ of $^{56}$Ni, and the synthetic multi-color
light curves  show good agreement with those observed
for normal SNe Ia. Due to the small number of such massive systems available, this model may only contribute a small fraction to the observed population of normal SNe Ia. Future studies
are needed to explore the parameter space of different WD masses and
mass ratios in this scenario for normal SNe Ia, which is important in BPS studies.

\subsection{Potential progenitor models}
Besides the SD and DD models above, some variants of SD and DD models
have been proposed to explain the observed diversity of SNe~Ia, such
as the sub-Ch
mass model, the super-Ch mass model, the single star model, the delayed
dynamical instability model, the spin-up/spin-down model,  the
core-degenerate model, the model of the collisions between two WDs, and the model of WDs near black holes, etc.

\subsubsection{Sub-Chandrasekhar mass model}
In this model, a CO WD accumulates a $\sim$$0.15\,M_{\odot}$  He
layer with a total mass below the Ch mass (Nomoto 1982b; Woosley et al., 1986).
In order to achieve the central densities necessary to produce
iron-peak elements, the WD in this model needs a narrow mass
range of $\sim$$0.9-1.1\,M_{\odot}$. The He may ignite
off-center at the bottom of the He layer, resulting in an event
known as Edge Lit Detonation (or Indirect Double Detonation).
In this process, one detonation propagates outward through the He
layer, while an inward propagating pressure wave compresses the CO
core that ignites off-center, followed by an outward detonation
(e.g. Livne, 1990; H\"{o}flich and Khokhlov, 1996). It
is possible that sub-luminous 1991bg-like objects may be explained by
this model (Branch et al., 1995). Unfortunately, the sub-Ch mass
model has difficulties in matching the observed light curves and
spectroscopy of SNe Ia (H\"{o}flich and Khokhlov, 1996; Nugent et al., 1997),
likely owing to the thickness of the He layer.

Recently, Shen and Bildsten (2009) argued that, under some suitable
conditions, a detonation in the WD might be achieved for even lower
He layer masses than that in previous studies. By assuming that a
detonation is successfully triggered in the He layer, Fink et
al. (2010) claimed that the double detonations in sub-Ch mass WDs with
low-mass He layers can be a robust explosion, leading
to normal SN~Ia brightness. Recent studies involving the sub-Ch mass
WDs with subsequent nucleosynthesis and radiative transfer
calculations also indicate that the sub-Ch mass model could account
for the range of the observed SN~Ia brightness (Sim et al., 2010; Kromer
et al., 2010).  Additionally, BPS studies by Ruiter
et al. (2009) predicted that there are a sufficient number of
binaries with sub-Ch primary WDs to explain the observed birthrate
of SNe~Ia.  However, it must be noted that it is difficult for
the sub-Ch mass model to explain the similarities observed in
most SNe~Ia (e.g. Branch et al., 1995).

\subsubsection{Super-Chandrasekhar mass model}
The $^{\rm 56}$Ni mass deduced from some SN~Ia explosions strongly
suggests the existence of super-Ch mass progenitors. SN 2003fg
was observed to be 2.2 times over-luminous than a normal SN~Ia, and
the amount of $^{\rm 56}$Ni was inferred to be $1.3\,M_{\odot}$, which
requires a super-Ch mass WD explosion ($\sim$$2.1\,M_{\odot}$;
Howell et al., 2006). Following the discovery of SN 2003fg,
three 2003fg-like events were also discovered, i.e. SN 2006gz
(Hicken et al., 2007), SN 2007if (Scalzo et al., 2010; Yuan et al.,
2010), and SN 2009dc (Yamanaka et al., 2009; Tanaka et al., 2010;
Silverman et al., 2011). These super-luminous SNe Ia  may raise the possibility that more
than one progenitor model may lead to SNe Ia.

It is usually assumed that these super-luminous SNe~Ia are from the mergers of
double WD systems, where the total mass of the DD systems is over the Ch
mass. Meanwhile, a super-Ch WD may be also produced by a SD system,
where the massive WD is supported by its rapid rotation, e.g. Maeda and Iwamoto (2009) claimed that the
properties of SN 2003fg may be consistent with the aspherical
explosion of a super-Ch WD, which is supported by its rapid rotation.
Yoon and Langer (2004) argued that WDs  can rotate differentially for
high mass-accretion rates of $3.0\times10^{-7}\,M_{\odot}\,\rm yr^{-1}$.
By adopting the results of Yoon and Langer (2004), Chen
and Li (2009) calculated the evolution of close binaries consisting
of a CO WD and a MS star, and obtained the initial parameter space for
super-Ch mass SN~Ia progenitors. Within this parameter space,
Meng et al. (2011) estimated that the upper limit of
the contribution rate of these super-luminous SNe~Ia to all SN Ia is
less than 0.3\%. Hachisu et al. (2012) recently made
a comprehensive study of these super-luminous SNe~Ia via the WD + MS
channel, and suggested that these SNe Ia are born in a low metallicity
environment as more massive initial CO WDs are required in this model.
Meanwhile, Liu et al. (2010) also studied the He star
donor channel to the formation of super-luminous SNe Ia by
considering the effects of rapid differential rotation on the
accreting WD.

Aside from the differential rotation, a super-Ch WD may also be
supported by the WDs with strong magnetic fields due to the lifting effect.
It has been found that $\sim$10\% of WDs have magnetic fields stronger than
1\,MG (Liebert et al., 2003, 2005; Wickramasinghe and Ferrario, 2005).
The mean mass of these magnetic WDs is $\sim$$0.93\,M_{\odot}$, compared with the mean
mass of all WDs that is $\sim$$0.56\,M_{\odot}$ (e.g. Parthasarathy
et al., 2007). Thus, the magnetic WDs are more easily
to reach the Ch mass limit by accretion.
The magnetic field may also affect some properties of SD progenitor
systems, e.g. the mass-transfer rate, the critical mass-accretion rate and
the thermonuclear reaction rate, etc. However, these effects are still unclear. Further studies are thus needed.

\subsubsection{Single star model}
Single star progenitor models have been considered by Iben and
Renzini (1983) and Tout et al. (2008). In the absence of mass-loss,
single massive star less than about 7\,$M_{\odot}$ will develop a
degenerate CO-core when the star evolves to the AGB stage. The
mass growing rate of the CO-core is controlled by the rate of the
double shell burning. If the CO-core can grow to the Ch mass, it will
produce a SN~Ia. Under certain conditions,
Tout et al. (2008) claimed that
carbon can ignite at the center of the CO-core and the subsequent
explosion would appear as a SN~Ia. These single star progenitors are
likely to be over 2\,$M_{\odot}$, so this kind of SNe~Ia should be
associated with younger galaxies with recent star formation.
The single star model was also proposed to explain the strongly
circumstellar-interacting SN 2002ic (Hamuy et al., 2003).

An important theoretical argument for this model is that the
H-rich envelope in AGB star may be lost in a superwind before
the CO-core grows to the Ch mass, based on the envelope ejection
criteria by Han et al. (1994) and Meng et
al. (2008). Another problem for this
model is that there should be far more SNe~Ia than observed
if a single star can naturally experience thermonuclear
explosion.

\subsubsection{Delayed dynamical instability model}

This model is a variant of the WD + MS channel, which
requires that the donor star is initially a
relatively  massive MS star ($\sim$3\,$M_{\odot}$) and that the
system has experienced a delayed dynamical instability, resulting in a
large amount of mass-loss from the system in the last a few
$10^{4}$\,yr before SN explosion (Han and Podsiadlowski, 2006). The delayed dynamical instability
model can reproduce the inferred H-rich circumstellar environment,
most likely with a disc-like geometry.
Han and Podsiadlowski (2006) claimed that the unusual properties of
SN 2002ic can be understood by the delayed dynamical instability
model. Observationally, this model
seems to be consistent with SN 2005gj (another 2002ic-like object) found
by Nearby Supernova Factory observations (Aldering et al., 2006).

However, in order for this model to be feasible, it requires a
larger mass-accretion efficiency onto the WD than is assumed in present
parametrizations. Based on a detailed BPS simulation, Han and
Podsiadlowski (2006) estimated that not more than 1\% SNe~Ia
should belong to this subclass of SNe~Ia. Since this model requires
an intermediate-mass secondary star, these SNe Ia should only be found
in stellar populations with relatively recent star formation (e.g.
with the last $\sim$$3\times10^{8}\,{\rm yr}$).

\subsubsection{Spin-up/spin-down model}

In the SD model, since the continued accretion of angular momentum can prevent
the explosion of a WD, Justham (2011) recently argued that it may
be natural for the mass donor stars in the SD model to exhaust their
envelopes and shrink rapidly before SN explosion, which may explain
the lack of H or He in the spectra of SNe~Ia, often seen as
troublesome for the SD progenitor model. Di Stefano et al. (2011) also
suggested that the CO WD is likely to achieve fast spin periods as
the accreted mass carries angular momentum, which can increase the
critical mass, $M_{\rm cr}$, needed for SN explosion. When the
$M_{\rm cr}$ is higher than the maximum mass obtained by the WD, the
WD must spin down before it explodes. This leads to a delay between
the time at which the WD has completed its epoch of mass gain and
the time of the SN explosion. However, the spin-down time is still
unclear, which may have a large range from
$<$1\,Myr to $>$1\,Gyr (Lindblom, 1999; Yoon and Langer,
2005).
The spin-down time may be important for the formation of the SNe Ia with long delay times.

The spin-up/spin-down model may provide a route to explain the
similarities and the diversity observed in SNe~Ia. However, the
birthrates, the delay times
and the distributions of SN~Ia explosion masses are still
uncertain in this model. A detailed BPS studies are needed for this.

\subsubsection{Core-degenerate model}
Kashi and Soker (2011) recently investigated some possible outcomes of
double WD mergers, in which these two components are made of CO. Most
simulations and calculations of  double WD mergers assume that a merger
occurs a long time after the CE ejection, when these two WDs are already
cold. In this model, Kashi and Soker (2011) proposed that, a
merger occurs within the final stages of the CE, whereas the CO-core is
still hot. The merged hot core is supported by rotation until it slows
down through the magnetic dipole radiation, and finally explodes.
Kashi and Soker (2011) named this as the core-degenerate model, and
claimed that this is another scenario to form a massive WD with
super-Ch mass that might explode as a super-luminous SN~Ia (see also
Ilkov and Soker, 2012).  A BPS study is required
to determine the birthrate and delay time of this model, which are then
compared with observations.

\subsubsection{Collisions of two WDs}
The WD number densities in globular clusters allow $\sim$10$-$100 times
collisions between two WDs per year, and the observations of globular clusters in
the nearby S0 galaxy NGC 7457 have detected a likely remnant of
SNe~Ia (Chomiuk et al., 2008). In this context, Raskin et al.
(2009) explored collisions between two WDs as a way for
producing SNe~Ia. They carried out simulations of the collisions
between two WDs ($\sim$$0.6\,M_{\odot}$) at various impact
parameters (the vertical separation of the centers of the WDs).
By taking impact parameters less than half of the WD radius before collision,
they claimed that the SN explosions induced by such collisions can produce
$\sim$$0.4\,M_{\odot}$ of $^{56}$Ni, making such objects potential
candidates for sub-luminous SN~Ia events. In a further study,
Raskin et al. (2010) argued that two WD collisions could also realize
super-Ch mass WD explosions (see also Rosswog et al., 2009a). However,
this model predicts a very aspherical explosion, inconsistent with
the small continuum polarization level in one of the observed
super-luminous SNe~Ia (i.e. SN 2009dc; see Tanaka et al., 2010).
We note that collisions between two WDs are likely to happen
in the dense environments of globular clusters, however the
expected of which is still less frequent than that of the
double WD mergers.

\subsubsection{WDs near black holes}
Wilson and Mathews (2004) proposed a new mechanism for producing
SNe~Ia, in which relativistic terms enhance the self-gravity of a CO
WD when it passes near a black hole. They suggested that this
relativistic compression can cause the central density of the WD to
exceed the threshold for pycnonuclear reactions so that a
thermonuclear runaway occurs. Dearborn et al. (2005) speculated that
this mechanism might explain the observed `mixed-morphology' of the Sgr A
East SN remnant in the Galactic center. For more studies of this
mechanism see Rosswog et al. (2008, 2009b). Due to the expected low rate of a
WD passing near a black hole, the expected SN~Ia
birthrate from this mechanism should be significantly lower than
that from normal SNe~Ia.

\section{Observational constraints }
Many observational results can be used to
constrain the SN Ia progenitor models, e.g. the properties of SN~Ia
host galaxies, the birthrates and delay times of SNe~Ia, the
candidate progenitors of SNe~Ia, the surviving companion stars of SNe~Ia,
the stripped mass of companions due to SN explosion, the signatures of gas
outflows from some SN~Ia progenitor systems, the wind-blown cavity in SN remnant,
the early optical and UV emission of SNe Ia, the early radio and X-ray emission of SNe Ia, and the pre-explosion images and spectropolarimetry of SNe Ia, etc.

\subsection{SN~Ia host galaxies}
There are some observational clues from the galaxies that host SNe
Ia. SNe~Ia have been known to occur both in young and old stellar
populations (e.g. Branch and van den Bergh, 1993), which implies
that there is a time delay between the star formation and the
SN explosion, ranging from much less than 1\,Gyr to at
least several Gyr. In addition, SNe~Ia in old population tend to be
less luminous, and the most luminous SNe~Ia appear to prefer young
populations with recent star formation (Hamuy et al., 1996; Wang et al. 2008a).
This indicates that the age of SNe Ia is
an important parameter controlling at least part of SN~Ia diversity.
It was also established that super-luminous SNe Ia preferably
occur in relatively metal poor environments with
low-mass host galaxies, whereas sub-luminous SNe Ia occur
in non star-forming host galaxies with large stellar
masses, such as elliptical galaxies (Neill et al., 2009; Taubenberger et al., 2011).

The observational homogeneity of SNe~Ia implies that a single
progenitor system may produce most or all SNe~Ia. However, evidence for
some observational diversity among SNe~Ia, as well as evidence that
SNe~Ia can be produced by stellar populations that have a wide
range of ages, raises the possibility that a variety of progenitor
systems may be contributing.

\subsection{Birthrates of SNe~Ia}

\begin{figure}[tb]
\includegraphics[width=6.0cm,angle=270]{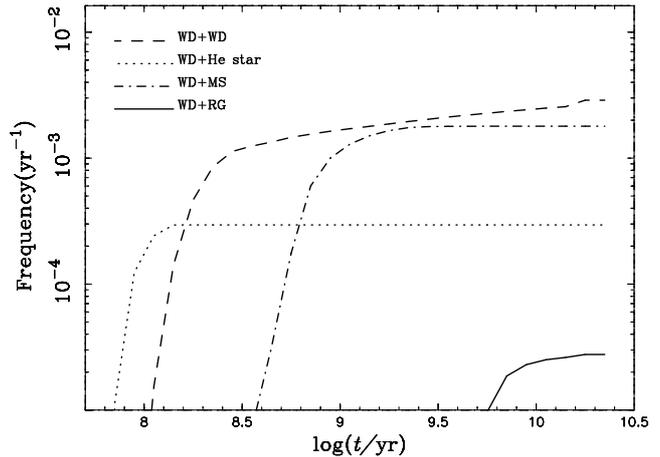}
 \caption{Evolution of Galactic SN~Ia birthrates for a constant star
formation rate ($Z=0.02$, ${\rm SFR}=5\,M_{\rm\odot}{\rm yr}^{-1}$).
(From Wang et al., 2010b)}
\end{figure}

The observed SN~Ia birthrate in our Galaxy is
$\sim$3$\times10^{-3}\,{\rm yr}^{-1}$ (Cappellaro and Turatto,
1997), which can be used to constrain the progenitor models of
SNe~Ia. Based on a detailed BPS study, Wang et al. (2010b)
systematically investigated Galactic SN Ia birthrates for the SD
and DD models, where the SD model includes the WD + MS, WD + RG and
WD + He star channels (see Fig. 4). They found that the Galactic SN~Ia
birthrate from the DD model is up to $2.9\times10^{-3}\,{\rm yr}^{-1}$
by assuming that SNe~Ia arise from the merging of two CO WDs
that have a combined mass larger than or equal to the Ch mass, which is
consistent with the birthrate inferred from observations,
whereas the total birthrates from the SD models can only account for
about 2/3 of the observations, in which the birthrate from the WD + MS
channel is up to $1.8\times10^{-3}\,{\rm yr}^{-1}$, the WD + RG channel is
up to $3\times10^{-5}\,{\rm yr}^{-1}$ and the WD + He star channel is
up to $0.3\times10^{-3}\,{\rm yr}^{-1}$. The Galactic SN~Ia
birthrate from the WD + RG channel is too low to be compared with
that of observations, i.e. SNe~Ia from this channel may be rare. However,
further studies on this channel are necessary, since this channel may
explain some SNe~Ia with long delay times. In addition, it has been
suggested that both recurrent novae, i.e. RS Oph and T CrB, are probable
SN~Ia progenitors and belong to the WD + RG channel (e.g.
Belczy$\acute{\rm n}$ski and Mikolajewska, 1998; Hachisu et al.,
1999b; Sokoloski et al., 2006; Hachisu et al., 2007; Patat et al., 2007a, 2011).
For other arguments in favour of the WD + RG channel see Sects. 4.2 and 4.3.

The SN Ia birthrate in galaxies is the convolution of
the delay time distributions (DTDs) with the star formation
history (SFH):
\begin{equation}
\nu(t)=\int^t_0 SFR(t-t')DTDs(t')dt',
\end{equation}
where $SFR$ is the star formation rate, and $t'$ is the delay
time of a SN Ia. Due to a constant $SFR$  adopted here, the SN
Ia birthrate $\nu(t)$ is only related to the $DTDs$, which can be
expressed by
\begin{equation}
DTDs(t)=\left\{
\begin{array}{lc}
0, & t<{t_1},\\
DTDs'(t) , &   {t_1} \leq t \leq{t_2},\\
0, & t>{t_2},\\
\end{array}\right.
\end{equation}
where ${t_1}$ and ${t_2}$ are the minimum and maximum delay
times of SNe Ia, respectively, and the $DTDs'$ is the distribution of
the delay times between ${t_1}$ and ${t_2}$. If $t$ is
larger than the ${t_2}$, equation (3) can be written as
\begin{equation}
\nu(t)={\rm SFR}\int^{t_2}_{t_1}DTDs'(t')dt'={\rm constant}.
\end{equation}
Therefore, the SN Ia birthrates shown in Fig. 4 seem to be
completely flat after the first rise.

\subsection{Delay time distributions}
\begin{figure*}
\centerline{\epsfig{file=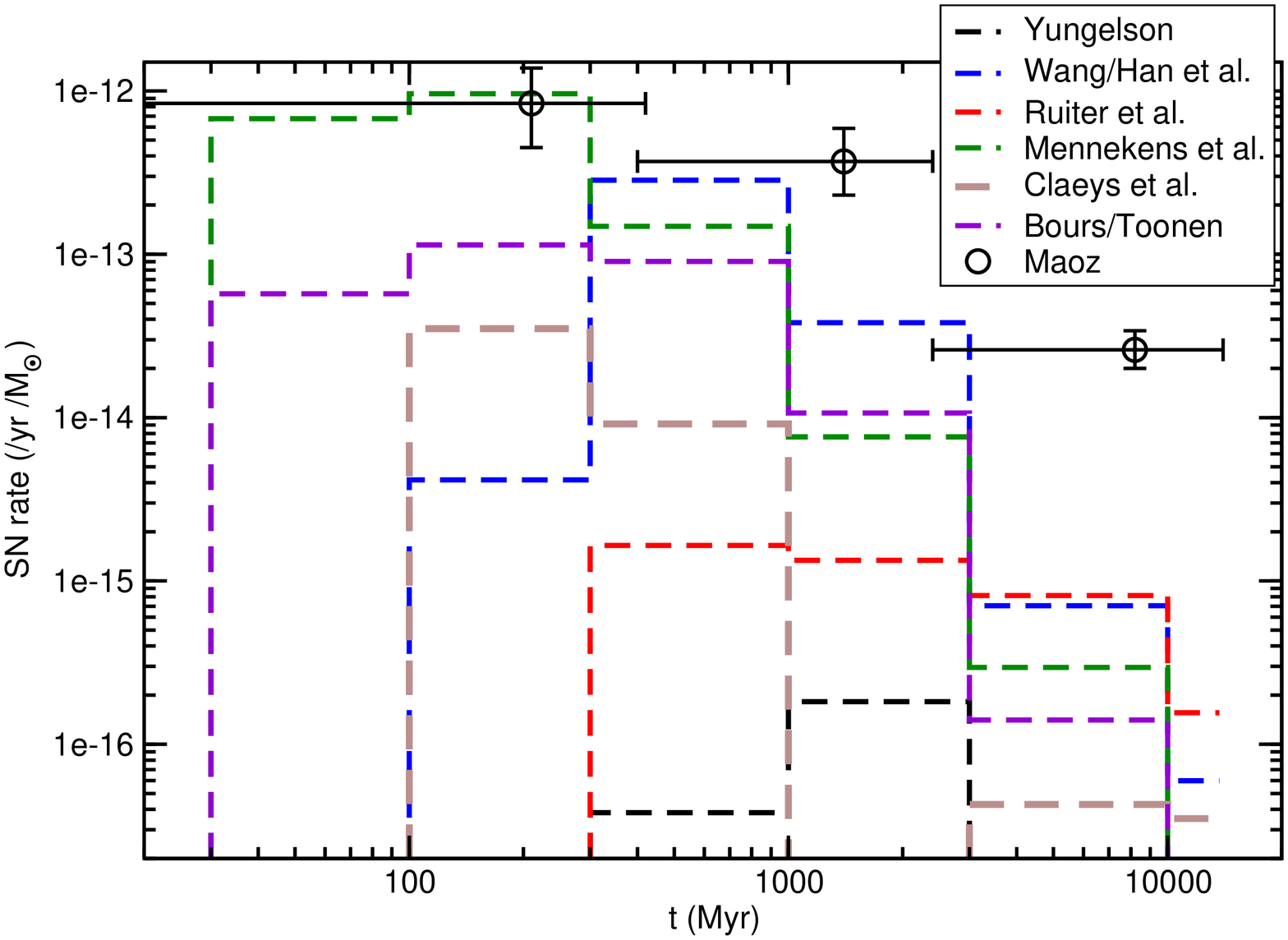,angle=270,width=9.5cm}\ \
\epsfig{file=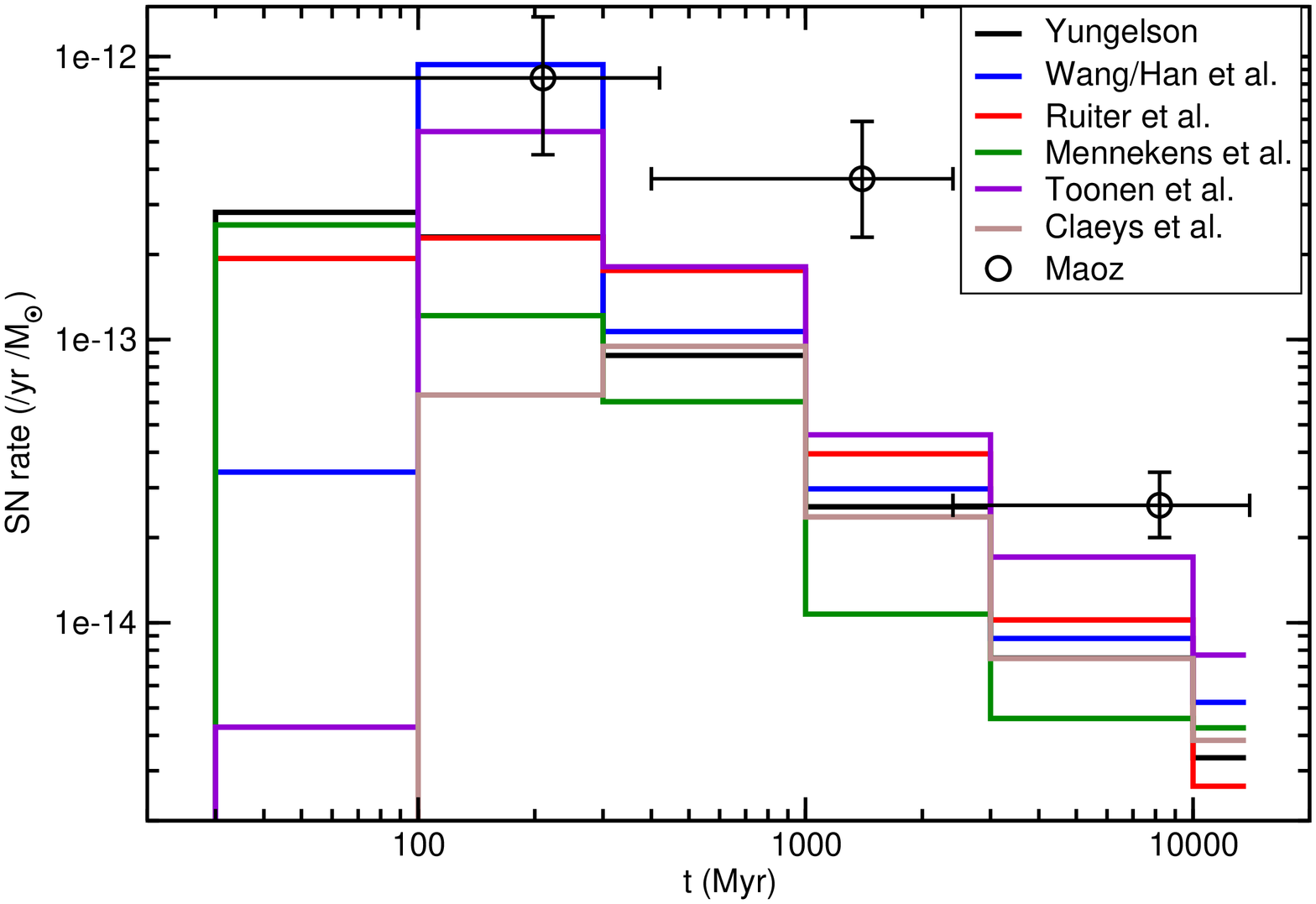,angle=270,width=9.5cm}} \caption{Rescaled delay
time distributions from different BPS research groups, for the SD (left
panel) and DD (right panel) models, respectively. The data
points are taken from Maoz et al. (2011). (From Nelemans et al., 2011)}
\end{figure*}

The delay times of SNe Ia are defined as the time interval between the star formation and SN explosion.
The various progenitor models of SNe~Ia can be examined by comparing
the delay time distributions (DTDs) expected from a progenitor model
with that of observations. Many works involve the observational DTDs
(e.g. Scannapieco and Bildsten, 2005; Mannucci et al., 2006, 2008;
F\"{o}rster et al., 2006; Aubourg et al., 2008; Botticella et al.,
2008; Totani et al., 2008; Schawinski, 2009; Maoz et al., 2011). In recent years, three important
observational results for SNe~Ia have been proposed, i.e. the strong
enhancement of the SN~Ia birthrate in radio-loud early-type
galaxies, the strong dependence of the SN~Ia birthrate on the colors
of the host galaxies, and the evolution of the SN~Ia birthrate with
redshift. Mannucci et al. (2006) claimed that these observational
results can be best matched by a bimodal DTD, in which about half of
SNe~Ia explode soon after starburst with a delay time less than
100\,Myr, whereas others have a much wider distribution with a delay
time $\sim$3\,Gyr. In a further study, Mannucci et al. (2008)
suggested that 10\% (weak bimodality) to 50\% (strong bimodality) of
all SNe~Ia belong to the young SNe~Ia. The existence of the young SN~Ia
population has also been confirmed by many other observations
(e.g. Aubourg et al., 2008; Cooper et al., 2009; Thomson and Chary, 2011),
although with a wide range in defining the delay times of
the young population.

Maoz et al. (2011) presented a new method to recover the DTD, which
can avoid some loss of information. In this method, the star
formation history of
every individual galaxy, or even every galaxy subunit, is convolved
with a trial universal DTD, and the resulted current SN~Ia
birthrate is compared to the number of SNe~Ia the galaxy hosted in
their survey. They reported that a significant detection of both a prompt
SN~Ia component, that explodes within 420\,Myr of star formation,
and a delayed SNe~Ia with population that explodes after 2.4\,Gyr.
Recently, a number of DTD measurements show that the DTD of SNe
Ia follows the power-law form of $t^{-1}$ (Maoz and Mannucci, 2012).
The power-law form is even different from the strong bimodal DTD
suggested by Mannucci et al. (2006), which might indicate that the
two-component model is an insufficient description for the
observational data. We also note that there are many uncertainties in the observed DTDs, which are dominated by the
uncertainties in galactic stellar populations and star formation histories (Maoz and Mannucci, 2012).

Many BPS groups work on the theoretical DTDs of SNe~Ia (e.g. Yungelson and
Livio, 2000; Nelemans et al., 2001; Han and Podsiadlowski, 2004;
Wang et al., 2009b, 2010a,b; Ruiter et al., 2009, 2011; Meng and Yang, 2010a;
Mennekens et al., 2010; Yu and Jeffery, 2011; Claeys and Pols, 2011).
Other theoretical DTDs of SNe Ia have been based on
physically motivated mathematical parameterizations
(e.g. Greggio and Renzini, 1983; Madau et al., 1998; Greggio, 2005, 2010).
Recently, Nelemans
et al. (2011) collected data from different BPS groups and made
a comparison. They found that the DTDs of different research groups for the
DD model agree reasonably well, whereas the SD model have rather
different results (see Fig. 5). One of the main differences in the
results of the SD model is the mass-accretion efficiency with which the
accreted H is added onto the surface of the WD (Nelemans et
al., 2011). However, the treatment of the mass-accretion efficiency
cannot explain all the differences. Nelemans et al. are planning to
do that in a forthcoming paper. For the SD model, Nelemans et al.
(2011) only considered systems with H-rich donor stars, not including
the He-rich donor stars (Wang et al., 2009a). It is worth noting that the He star
donor channel can produce SNe~Ia effectively with short delay times (accounting
for 14\% of all SNe~Ia in SD model; Wang et al., 2010b), which
constitutes the weak bimodality as suggested by Mannucci et al.
(2008).

Hachisu et al. (2008) recently investigated
new binary evolutionary models for SN~Ia progenitors, with  introducing the
mass-stripping effect on a massive MS companion star by
winds from a mass-accreting WD. This model can also provide a possible way for producing young SNe
Ia, but the model significantly depends on the
efficiency of the artificial mass-stripping effect. Additionally,
Chen and Li (2007) studied the WD + MS channel by
considering a circumbinary disk which extracts the orbital angular
momentum from the binary through tidal torques. This study also provides
a possible way to produce SNe~Ia with long delay times ($\sim$1$-$3\,Gyr).

\subsection{Candidate progenitors}
\subsubsection{Single-degenerate progenitors}
A number of WD binaries are known to be excellent candidates for SD
progenitors of SNe~Ia, e.g. U Sco, RS Oph and TCrB  (Parthasarathy et al., 2007).
All of these binaries contain WDs
which are already close to the Ch mass, where the latter two systems
are symbiotic binaries containing a giant companion star (see
Hachisu et al., 1999b). However, it is unclear whether these massive
WD is a CO or an O-Ne-Mg WD; the latter is thought to collapse by
forming a neutron star through electron capture on $^{24}$Mg rather
than experience a thermonuclear explosion (for more discussion see Sect. 4.2).
Meanwhile, there are also two massive WD + He star systems (HD 49798
with its WD companion and V445 Pup), which are good candidates of SN~Ia
progenitors.

HD 49798 is a H depleted subdwarf O6 star and also
a single-component spectroscopic binary with an orbital period of
1.548\,d (Thackeray, 1970), which contains an X-ray pulsating
companion star (RX J0648.0-4418; Israel et al., 1997). The X-ray
pulsating companion star is suggested to be a massive WD
(Bisscheroux et al., 1997). Based on the pulse time delays and the
inclination of the binary, constrained by the duration of the X-ray
eclipse, Mereghetti et al. (2009) recently derived the masses of
these two components. The corresponding masses are
1.50$\pm$0.05$\,M_{\odot}$ for HD 49798 and
1.28$\pm$0.05$\,M_{\odot}$ for the WD. According to a detailed
binary evolution model with the optically thick wind assumption,
Wang and Han (2010c) found that the massive
WD can increase its mass to the Ch mass after only a few $10^{4}$\,years. Thus,
HD 49798 with its WD companion is a likely candidate of a SN Ia
progenitor.

V445 Pup is the first, and so far only,
helium nova detected (Ashok and Banerjee, 2003; Kato and Hachisu, 2003).
The outburst of V445 Pup was discovered on 30 December 2000
by Kanatsu (Kato et al., 2000). After that time, a dense dust shell was formed in the ejecta of the outburst, and the star became a strong
infrared source, resulting in the star's fading below 20\,magnitudes in the $V$-band (Goranskij et al., 2010).
From 2003 to 2009, $BVR$
observations by Goranskij et al. (2010) suggest that the dust absorption minimum finished in
2004, and the remnant reappeared at the level of 18.5\,magnitudes in the $V$-band.
Goranskij et al. (2010) reported that
the most probable orbital period of the binary system is $\sim$0.65\,day.
Based on the optically thick wind theory, Kato et al. (2008)
presented a free-free emission dominated light curve model of V445
Pup. The light curve fitting in their study shows that the mass of
the WD is more than $1.35\,M_{\odot}$, and half of the accreted
matter remains on the WD, leading to the mass increase of the WD.
In addition, the massive WD is a CO WD instead of an O-Ne-Mg WD, since no indication
of neon was observed in the nebula-phase spectrum (Woudt
and Steeghs, 2005). Therefore, V445 Pup is a strong
candidate of a SN~Ia progenitor (e.g. Kato et al., 2008; Woudt et al., 2009).

\subsubsection{Double-degenerate progenitors}
Several systematic searches for double WD systems have been made. The
largest survey for this is SPY (ESO SN~Ia Progenitor Survey;
Napiwotzki et al., 2004; Nelemans et al., 2005; Geier et al., 2007),
which aims at finding double WD systems as candidates of SN~Ia
progenitors. The only likely SN~Ia progenitor in this sample is not a
double WD system, but the WD + sdB binary KPD 1930+2752 (Maxted et al., 2000). The
orbital period of this binary is 2.283\,h, the mass of the sdB star
is $\sim$$0.55\,M_{\odot}$, and the mass of the WD is $\sim$$0.97\,M_{\odot}$.
The total mass ($\sim$$1.52\,M_{\odot}$) and the merging time ($<$0.2\,Gyr) of the binary
indicate that it is a good candidate of a SN~Ia progenitor
(Geier et al., 2007).

Recently, some other
double WD systems have also been found, which may have the total mass close to the Ch mass, and
possibly merge in the Hubble-time. These include a binary WD 2020-425 with $P_{\rm orb}\sim0.3$\,day,
$M_{\rm 1}+M_{\rm 2}=1.348\pm0.045\,M_{\odot}$ (Napiwotzki et al., 2007),
V458 Vulpeculae with $P_{\rm orb}\sim0.068$\,day,
$M_{\rm 1}\sim0.6\,M_{\odot}$, $M_{\rm 2}>1.0\,M_{\odot}$ (Rodr\'{\i}guez-Gil et al., 2010),
a close binary star SBS 1150+599A (double-degenerate nucleus of the planetary nebula TS 01) with
$P_{\rm orb}\sim0.163$\,day,
$M_{\rm 1}=0.54\pm0.02\,M_{\odot}$, $M_{\rm 2}\sim0.86\,M_{\odot}$ (Tovmassian et al., 2010),
and  GD687 that will evolve into a double WD system and merge to form a rare supermassive WD with the total mass
at least $1\,M_{\odot}$ (Geier et al., 2010). There are also some ongoing projects searching for double WD systems,
e.g. the SWARMS survey by Badenes et al. (2009b) which is searching for compact WD
binaries based on the spectroscopic catalog of the Sloan Digital Sky Survey.

\subsection{Surviving companion stars}
A SN Ia explosion following the merger of two WDs will leave no compact remnant behind,
whereas the companion star in the SD model will survive after a SN explosion and
potentially be identifiable by virtue of its anomalous properties. Thus,
one way to distinguish between the SD and DD models is to look at
the center of a known SN Ia remnant to see whether any surviving companion star is present.
A surviving companion star in the SD model would evolve to a WD
finally, and Hansen (2003) suggested that the SD model could
potentially explain the properties of halo WDs (e.g. their space
density and ages). Note that, there has been no conclusive proof yet
that any individual object is the surviving companion star of a
SN~Ia. It will be a promising method to test SN~Ia progenitor models
by identifying their surviving companions.

Han (2008) obtained many properties of the surviving companion stars of
SNe~Ia with intermediate delay times (100\,Myr$-$1\,Gyr) from the WD + MS channel, which
are runaway stars moving away from the center of SN remnants. Wang
and Han (2009) studied the properties of the companion stars of the
SNe~Ia with short delay times ($<$100\,Myr) from the He star donor
channel, which are related to hypervelocity He stars (for more
discussion see Sect. 4.5; also see Justham et al., 2009).
Moreover, Wang and Han (2010d) recently
obtained the properties of the surviving companions of the SNe~Ia
with long delay times ($>$1\,Gyr) from the WD + MS and WD + RG
channels, providing a possible way to explain the formation of
the population of single low-mass He WDs ($<$0.45$\,M_{\odot}$; for
more discussion see Sect. 4.4; also see Justham et al., 2009). The properties of the surviving
companion stars (e.g. the masses, the spatial velocities, the
effective temperatures, the luminosities and the surface gravities,
etc) can be verified by future observations.

Tycho G was taken as the surviving companion of Tycho's SN by
Ruiz-Lapuente et al. (2004). It has a space velocity of $136\,{\rm
km/s}$, more than three times the mean velocity of the stars in the
vicinity. Its surface gravity is $\log\, (g/{\rm cm}\, {\rm
s}^{-2})=3.5\pm 0.5$, whereas the effective temperature is $T_{\rm
eff}=5750\pm 250 {\rm K}$ (Ruiz-Lapuente et al., 2004).
These parameters are compatible with the
properties of SN~Ia surviving companions from the SD model (e.g. Han, 2008; Wang
and Han, 2010d). However, Fuhrmann (2005) argued that
Tycho G might be a Milky way thick-disk star that is coincidentally
passing the vicinity of the remnant of Tycho's SN. Ihara et al.
(2007) also argued that Tycho G may not be the companion star of
Tycho's SN, since this star does not show any special properties in
its spectrum; the surviving companions of SNe Ia would be
contaminated by SN ejecta and show some special
characteristics.\footnote{Pan et al. (2012) studied the impact of SN Ia
ejecta on MS, RG and He star companions with the FLASH
code. They
quantified the amount of contamination on the companion star by
the SN ejecta in their simulations, which might help to
identify a companion star even a long time after the SN explosion.}
Recently, Gonz$\acute{\rm
a}$lez-Hern$\acute{\rm a}$ndez et al. (2009) presented some evidence
that Tycho G may be enriched in $^{56}$Ni, which could be the result of
pollution of the atmosphere with the SN ejecta.

By assuming that the companion star in the SD model is
co-rotating with the binary orbit at the moment of the SN explosion, the predicted rotational velocity
of Tycho G is $\sim$$100\,\rm km/s$  (e.g. Wang and Han, 2010d). However, the rapid rotation
predicted by the SD model is not observed in Tycho G ($7.5\pm2\,\rm km/s$; Kerzendorf et al., 2009).
This does not yet rule out that this star is the surviving companion. The inferred  slow
rotation of Tycho G may be related to the angular momentum loss induced
by the rapid expansion of its outer shell. Recently, Pan et al. (2012) claimed
that the post-impact companion star loses about half of its initial angular momentum for
Tycho G, with the rotational velocity decreasing to a quarter
of its initial rotational velocity, $\sim$$37\,\rm km/s$,
which is closer to the observed rotational velocity ($7.5\pm2\,\rm km/s$).
Therefore, whether Tycho G is
the surviving companion of Tycho's SN is still quite debatable. The
confliction might be conquered by studying the interaction between the
SN ejecta and the rotating companion star.

We also note that Lu et al. (2011) recently claimed that
the angle between the direction of the non-thermal X-ray arc in Tycho's
SNR to the explosive center and the proper motion velocity of Tycho G
is well consistent with the theoretical predictions and simulations. This
supports Tycho G as the surviving companion of Tycho's SN.
Lu et al. (2011) also estimated the parameters of
the binary system before the SN explosion, which is useful
for constraining progenitor models of SNe Ia.

By investigating archival Hubble Space Telescope deep images,
Schaefer and Pagnotta (2012) recently reported that the central
region of SNR 0509-67.5 (the site of a 1991T-like SN Ia explosion
that occurred $\sim$400 years ago)
in the Large Magellanic Cloud contains no surviving companion star.
Thus, they argued that the progenitor of this particular SN Ia is a double WD system.
In a subsequent work, Edwards et al. (2012) used the same method as in
Schaefer and Pagnotta (2012) on SNR 0519-69.0, which is a normal SN Ia
remnant in the Large Magellanic Cloud with an age of 600$\pm$200 years, and
found that the 99.73\% error circle contains no post-MS stars for SNR 0519-69.0.
Thus, Edwards et al. (2012) claimed to
rule out the symbiotic, recurrent nova, He star and spin-up/spin-down
models for this particular SN. They argued that
SNR 0519-69.0 might be formed from either a supersoft channel or a double WD
merge. We note that, based on very short maximum spin-down times,
Edwards et al. (2012) excluded the spin-up/spin-down
model. However, if the spin-down time
is much longer, the results in Edwards et al. (2012) might be different.

\subsection{Stripped mass of companions}
In the SD model, SN explosion will strip some mass of its
non-degenerate companion star. By using
two-dimensional Eulerian hydrodynamics simulations, Marietta et al. (2000) examined the interaction of SN
ejecta with a MS star, a subgiant star and a RG star. They claimed that the MS and subgiant
companions lose $\sim$10$-$20\% of their mass after the SN explosion, and the RG companion loses about
96\%$-$98\% of its envelope. In this process, these stripped material is mixed with the SN ejecta.
Since these stripped material is likely to be dominated by H,
this should then lead to easily detectable H emission lines in the SN
nebular phase. Unfortunately, no H has ever been
detected in a normal SN~Ia. The most
recently observational upper limits on the amount of H detected are
$\sim$$0.01\,M_{\odot}$ (Leonard, 2007),\footnote{Leonard
(2007) obtained deep spectroscopy in the late nebular phase of
two well observed SNe~Ia (SN 2005am and SN 2005cf), in search of the trace amounts of H and He that
would be expected from the SD model.} which may provide a strong
constraint on the progenitor model of SNe~Ia. Additionally, based on the
properties of the X-ray arc inside the Tycho's SNR, Lu et al. (2011)
also obtained a low stripped mass ($\leq$$0.0083\,M_{\odot}$), consistent
with that from Leonard (2007).
These observational
limits are inconsistent with Marietta's predictions.

Meng et al. (2007) used a simple analytical method to calculate
the amount of the stripped masses. They obtained a lower limit of
$0.035\,M_{\odot}$ for the stripped mass, but their analytic method used
oversimplified physics of the interaction between SN Ia ejecta
and a companion star. Recently, many updated studies involve the effects of SN
explosion on the companion star. However, more realistic stellar
models for the companion star do not show  stripped mass
as small as that close to the Leonard's observational limits, i.e. they do
not resolve the conflict between the theory and the observations
(Pakmor et al., 2008; Pan et al., 2010, 2012; Liu et
al., 2012b). Thus, the high stripped mass from simulations may bring some problems for
the SD model. The spin-up/spin-down model may explain
the lack of H or He in SNe Ia (Justham, 2011; Di Stefano et
al., 2011; Hachisu et al., 2012). In addition, the mixture degree between the SN ejecta and
the stripped material may also influence the detection of H or He lines
in the nebular spectra of SNe Ia.

\subsection{Circumstellar material after SN explosion}
In the SD model, non-accreted material blown away from the binary
system before SN explosion should remain as circumstellar matter
(CSM). Thus, the detection of CSM in SN~Ia early
spectra would support the SD model. Patat et al. (2007a) found some
direct evidence on CSM in a normal SN~Ia, i.e. SN 2006X, which was also exceptional
in its high ejecta velocity and high reddening (Wang et al., 2008b).
Patat et al. (2007a) have
observed a variation of Na~I doublet lines immediately after the SN
explosion, which is interpreted as arising from the ionization and
subsequent recombination of Na in CSM. This strongly favours a SD
progenitor for this SN. Patat et al. (2007a) suggested that the
narrow lines may be explained by a recurrent nova.
The time-variable Na~I doublet absorption features
are also found in SN 1999cl (Blondin et al., 2009) and SN 2007le (Simon et al., 2009).
Patat et al. (2007a) argued that the CSM may be common
in all SNe~Ia, although there exists variation in its detect
ability because of viewing angle effects. However, in a subsequent work,
Patat et al. (2007b) did not find the same spectral features in SN 2000cx as they did with SN 2006X,
which indicates that there might be multiple SD progenitor models.
Meanwhile, the derivation of smaller
absorption ratio $R_{\rm V}$ (the ratio of the total to selective
absorption by dust) perhaps also suggests the presence of CSM dust around a
subclass of SNe~Ia (Wang et al., 2009c).

More encouragingly, Sternberg et al. (2011) studied the
velocity structure of absorbing material along the line of sight to
35 SNe~Ia in nearby spiral galaxies via Na~I doublet absorption features. They found a strong statistical
preference for blue shifted structures, which are likely signatures
of gas outflows from the SN Ia progenitor systems. They concluded
that many SNe~Ia in nearby spiral galaxies may originate in SD
systems, and estimated that at least 20\% of SNe~Ia that occur
in spiral galaxies are from the SD progenitors. Recently,
Foley et al. (2012) reported that SNe Ia with blue shifted structures
have higher ejecta velocities and redder colors at maximum brightness relative
to the rest of the SN Ia population, which
provides the link between the progenitor systems and properties of SN explosion.
This result adds additional confirmation that some SNe Ia are produced from the SD
model. However, Shen
et al. (2012) argued that such gas outflow signatures
could also be induced by winds and/or the
mass ejected during the coalescence in the double WDs.

\subsection{SN remnants}
SN remnants (SNRs) are beautiful astronomical objects that are also of
high scientific interest, since they provide direct insights into SN
progenitor models and explosion mechanisms. Recent studies by
Lu et al. (2011) suggested that the non-thermal X-ray arc in Tycho's
SNR is a result of interaction between the SN ejecta and the stripped
mass of the companion, strengthening the motivation of studying
the progenitor of a SN by studying its SNR. In addition, SNRs
may reveal the metallicity of SN progenitors (Badenes et al., 2008).

Circumstellar matter (CSM) is predicted by the SD model, which was
responsible for creating a low-density bubble (i.e. wind-blown cavity; Badenes et al., 2007). Its
modification on larger scales will become apparent during the SNR phase.
One of the obstacles the SD model faces is to search for this signatures from SNR
observations. Badenes et al. (2007) searched 7 young SN~Ia remnants
for the wind-blown cavities that would be expected in the
SD model. Unfortunately, in every case it appears that
the remnant is expanding into a constant density interstellar matter
(i.e. there is no wind-blown cavity in these SN remnants). However,
Williams et al. (2011) recently reported results from a
multi-wavelength analysis of the Galactic SN remnant RCW 86 (remnant
of SN 185 A.D.). From hydrodynamic simulations, the observed
characteristics of RCW 86 are successfully reproduced by an
off-center SN explosion in a low-density cavity carved by the
progenitor system (Williams et al., 2011). This makes RCW 86 the first known case of a SN~Ia
in a wind-blown cavity.

\subsection{Early optical and UV emission of SNe Ia}
The presence of a non-degenerate companion in the SD model
could leave an observable trace in the form of the optical and ultraviolet (UV) emission.
Kasen (2010) showed that the collision of the SN ejecta with its companion should
produce detectable optical and UV emission in the hours and days following the
SN explosion, which can be used to infer the radius of the companion. 
Thus, the early optical and UV observations of SN ejecta
can directly test progenitor models.
The optical and UV emission at early times forms mainly in
the outer shells of the SN ejecta, in which the unburned
outer layers of the WD  play an important role in shaping
the appearance of the spectrum. Kasen (2010) claimed that
these emission would be
observable only under favorable viewing angles, and its intensity
depends on the nature of the companion star.

Hayden et al. (2010) looked for this signal in the rising
portion of the $B$-band light curves of 108 SNe Ia from Sloan Digital Sky Survey, finding
no strong evidence of a shock signature in the data. They constrained
the companion in the SD model to be less than a
$6\,M_{\rm \odot}$ MS star, strongly disfavouring a RG star
undergoing RLOF. Recently,
Bianco et al. (2011) searched for the signature of a non-degenerate companion
star in three years of SN Legacy Survey data by generating synthetic
light curves accounting for the shock effects and comparing true and
synthetic time series with Kolmogorov-Smirnov tests. Based on the
constraining result that the shock effect is more prominent in
rest-frame $B$ than $V$ band (for details see Fig. 3 of Kasen, 2010),
Bianco et al. (2011) excluded a contribution of WD + RG binaries to SN Ia explosions.
However, a rather contradictory result for the shock effects was obtained by Ganeshalingam et al. (2011).

These shock signatures predicted in Kasen (2010)
are based on the assumption that the companion star fills its Roche
lobe at the moment of a SN explosion. However, if the binary separation is much larger than
the radius of the companion star, the solid angle subtended by the companion would be much smaller. Thus,
the shock effect would be lower. Justham (2011) and  Di Stefano et al.
(2011) argued that the donor star in the SD model may shrink
rapidly before the SN explosion, since it would exhaust its H-rich envelope during a long spin-down time of
the rapidly rotating WD until the SN explosion. In this condition, the companion star would be a smaller target for
the SN ejecta and produce a much smaller shock luminosity than the Roche lobe model considered in Kasen (2010) (see also Hachisu et al., 2012). Therefore, the early optical and UV emission of SN ejecta may be compatible with the SD model.

In recent optical and UV observations,  Wang et al. (2012) presented UV
and optical photometry and early time
spectra of four SNe Ia (SNe 2004dt, 2004ef,
2005M, and 2005cf) by using Hubble Space
Telescope. One SN Ia in their sample, SN 2004dt, displays a UV excess
(the spectra reveal an excess in the 2900$-$3500\,{\AA} wavelength range, compared with spectra of the other
SN Ia events). In their study, the comparison object SN 2006X may also
exhibit strong UV emission. The early UV emission may indicate the presence of a non-degenerate companion star in
SN Ia progenitor systems.

\subsection{Early radio and X-ray emission of SNe Ia}

Circumstellar matter (CSM) provides a medium with which the SN ejecta can interact and produce
radio synchrotron emission. Many authors have searched for early radio emission
from SNe Ia, but no detection has been made (Weiler et al., 1989; Eck et al., 1995, 2002).
Hancock et al. (2011) recently have used a stacking analysis of 46
archival Very Large Array observations by Panagia et al. (2006)
to set upper limits on the radio emission from SNe Ia in nearby galaxies.
They gave an upper limit on the SN Ia peak radio luminosity
of $1.2\times10^{25}\,{\rm erg\,s^{-1}\,Hz^{-1}}$ at 5\,GHz, which implies
an upper limit on the average companion stellar wind mass-loss rate of
$1.3\times10^{-7}\,M_{\odot}\,\rm yr^{-1}$ before a SN explosion.
Hancock et al. (2011) argued that
these limits challenge expectations if the SN ejecta were encountering a CSM from the SD model.

Aside from radio emission, the interaction of SN ejecta with the CSM can also produce
X-ray emission. SN shock would run into CSM and heat it to high enough temperatures
($\sim$$10^{6}-10^{9}$\,K), resulting in thermal X-rays (Chevalier, 1990).
Compared with radio emission, X-rays from SNe Ia result from a different
process and from different regions in
the shocked CSM. Thus, it is a completely independent method
to constrain progenitor model via detecting early X-ray emission of SNe Ia.
Russel and Immler (2012) recently considered 53 SNe Ia observed by the Swift X-Ray Telescope.
They gave an upper limit on the X-ray luminosity ($0.2-10$\,keV) of $1.7\times10^{38}\,{\rm erg\,s^{-1}}$,
which implies an upper limit on mass-loss rate of
$1.1\times10^{-6}\,M_{\odot}\,\rm yr^{-1}\times(\nu_{w})/(10\,km\,s^{-1})$, where
$\rm \nu_{w}$ is the wind speed for red supergiants that ranges from 5 to 25\,$\rm km\,s^{-1}$.
Russel and Immler (2012) claimed that these limits exclude massive or evolved
stars as the companions in progenitor systems of SNe Ia, but allow the
possibility of MS and WD as the companion.

According to the spin-up/spin-down model of SNe Ia suggested by Justham (2011) and Di Stefano et al. (2011),
there is a delay between the time at which the WD has completed its mass-accretion and
the time of the SN explosion. Since the matter ejected from the binary
system during the mass-transfer has a chance to become diffuse, the
SN explosion will occur in a medium with a density similar to that of
typical regions of the interstellar medium.
Therefore, the SD model may be compatible with the upper limits from SN Ia radio and X-ray detection.

\subsection{Pre-explosion images}
One of the methods to clarify SN~Ia progenitor models is to directly
detect the progenitor of a SN~Ia in pre-explosion images of the
position where the SN occurred. Voss and Nelemans (2008)
first studied the pre-explosion archival X-ray images at the
position of the recent SN 2007on, and considered that its progenitor
may be a WD + RG system. However, Roelofs et al. (2008)
did not detect any X-ray source in images taken six
weeks after SN 2007on's optical maximum and found an
offset between the SN and the measured X-ray source
position. Nelemans et al. (2008) also obtained an ambiguous answer.
Nielsen et al. (2011) recently derived the upper limits of the X-ray
luminosities from the locations of ten SNe Ia
in nearby galaxies ($<$25\,Mpc) before the explosions,
most above a few $10^{38}\,{\rm erg\,s^{-1}}$ (for details see
Fig. 1 of Nielsen et al., 2011), which indicates
that the progenitors of these SNe Ia were not bright supersoft X-ray sources
shortly before they exploded as SNe Ia. However, the upper
limits are not constraining enough to
rule out less bright supersoft X-ray progenitors (Nielsen et al., 2011).
Future observations may
shed light on the connection between SN Ia progenitors and X-ray
emission.

SN 2011fe occurred in M101 at a distance of 6.4\,Mpc is the second closest SN~Ia in the
digital imaging era,\footnote{The closest SN Ia in the
digital imaging era is SN 1986G that exploded in
NGC 5128 at a distance of $\sim$4\,Mpc (Frogel et al., 1987).} which was discovered by the Palomar Transient Factory
survey less than a day after its explosion (Nugent et al., 2011a), and quickly
followed up in many wavebands (Li et al., 2011; 
Nugent et al., 2011b; Smith et al., 2011; Tammann and Reindl, 2011;
Patat et al., 2011b; Liu et al., 2012; Horesh et al., 2012; Chomiuk et al., 2012;
Bloom et al., 2012; Brown et al., 2012a; Margutti et al., 2012).
Li et al. (2011) used extensive historical imaging obtained
at the location of SN 2011fe to constrain the visible-light luminosity of
the progenitor to be 10$-$100 times fainter than previous limits on
other SN~Ia progenitors. This result rules out luminous RG
stars and most He stars as the mass donor star of this SN progenitor. These observations
favour a scenario where the progenitor of SN 2011fe accreted material
either from WD, or via RLOF from a MS or subgiant companion. In a subsequent work,
Liu et al. (2012) also excluded its progenitor system with the most hottest photospheres
by constraining X-ray properties prior to the SN explosion.

Very recently, Horesh et al. (2012) set upper limits on both radio and X-ray
emission from SN 2011fe, excluding the presence of a circumstellar
matter from a giant donor star.
Based on deep radio observations, Chomiuk et al. (2012)
also excluded the presence of circumstellar matter.
By using early optical and UV observations of SN 2011fe,
Nugent et al. (2011b) excluded the presence of shock effects from SN
ejecta hitting a companion, and put a strict upper limit to the exploding star radius
($\leq$$0.1\,R_{\odot}$), thus providing a direct evidence
that the progenitor is a compact star. A recent study by Bloom et al. (2012) also ruled out a MS star as
the mass donor star and seem to favor a DD progenitor for SN 2011fe (also see Brown et al., 2012a).
We note that the spin-up/spin-down model potentially affects the conclusions above.

\subsection{Polarization of SNe Ia}
Spectropolarimetry provides a direct probe of early time SN
geometry, which is an important diagnostic tool for discriminating among SN Ia
progenitor systems and theories of SN explosion physics (see Livio and Pringle, 2011).
A hot young SN atmosphere is dominated by the electron scattering that is highly
polarizing. For an unresolved source with a spherical distribution
of scattering electrons, the directional components of the electric
vectors of the scattered photons counteract exactly, resulting in zero net linear polarization.
However, an incomplete cancelation will be derived from
any asymmetry in the distribution of the scattering electrons, or of
absorbing material overlying the electron-scattering atmosphere, which produces
a net polarization (Leonard and Filippenko, 2005).

SN asymmetry can therefore be measured via spectropolarimetry, since
asymmetric electron scattering leads to polarization vectors that do not cancel.
Most normal SNe Ia are found to be spherically symmetric (a rather low polarization, $\lesssim$0.3\%; Wang et al., 1996; Wang and Wheeler, 2008), but asymmetry has been detected at significant levels for a range of SN Ia subclasses, e.g. sub-luminous SNe Ia with a continuum polarization about 0.3\%$-$0.8\% (Howell et al., 2001), and
high-velocity (HV) SNe Ia with a high polarization about 2\%, the spectra of which around maximum light are characterized
by unusually broad and highly blueshifted absorption troughs
in many line features (Leonard et al., 2005).
Leonard et al. (2005) claimed that the following order
emerges in terms of increasing strength of line-polarization features: normal/over-luminous SNe Ia $<$ sub-luminous SNe Ia $<$ HV SNe Ia. They argued that the most convincing explanation for the linear  polarization of all objects is
partial obscuration of the photosphere by clumps of intermediate-mass elements forged in the SN explosion. For a review of SN Ia polarimetric studies see Wang and Wheeler (2008).

The explosion mechanism itself may produce asymmetry due to off-center explosion, and thus a polarization spectrum is expected (Plewa et al., 2004; Kasen and Plewa, 2005). Thus, it is possible to obtain insight into the SN explosion physics with spectropolarimetry. Meanwhile, the progenitor systems may also cause the asymmetry.
The SD model provides a natural way to produce the asymmetry. The
existence of a companion in the SD model may change the configuration of the SN ejecta
(e.g. a cone-shaped hole shadowed by the companion),
and thus a polarization spectrum is expected  (Marietta et al., 2000; Kasen et al., 2004; Meng and Yang, 2010b).\footnote{By using smoothed particle hydrodynamics simulations, Garc\'{\i}a-Senz et al. (2012) studied the
interaction of the hole, SN material and ambient medium. They concluded
that the hole could remain open in the SNR for hundreds of years, suggesting the hole could affect
its structure and evolution.}
In addition, the DD model may also
naturally result in an asymmetry of the distribution of SN
ejecta. One relevant mechanism is the rapid rotation of a WD before a
SN explosion, which leads to a change in the stellar
shape. Another is that there may be a thick accretion disc around the
CO WD, which may be an origin of asymmetry in the configuration
of the SN ejecta (e.g. Hillebrandt and Niemeyer, 2000).

Livio and Pringle (2011) argued that the nature of the correlation
between the polarization and the observed SN Ia properties can be used
to distinguish between the SD and DD models. As a specific example,
they considered possible correlations between the polarization and
the velocity gradient; a SN explosion
is viewed from one pole it is seen as a high velocity gradient
event at early phases with redshifts in late-time emission
lines, while if it is viewed from the other pole it is seen as
a low-velocity gradient event with blueshifts at late phases (Maeda et al, 2010).
In the SD model, it is expected that the velocity gradient is a
two-valued function of polarization, with the largest and smallest values
corresponding to essentially zero polarization. In the DD model,
it is expected that the observed SN properties (i.e. velocity gradient)
is a single-valued and monotonic function
of polarization. For details see Fig. 1 of Livio and Pringle (2011).

\section{Related objects}
There are some objects that may be related to the progenitors and surviving
companions of SNe~Ia in observations, e.g. supersoft X-ray
sources, cataclysmic variables, symbiotic systems,
single low-mass He WDs and hypervelocity He stars, etc.

\subsection{Supersoft X-ray sources}
Supersoft X-ray sources (SSSs) are one of the most promising
progenitor candidates of SNe~Ia. Binaries in which steady nuclear burning
takes place on the surface of the WDs have been  identified with bright
SSSs, discovered by the ROSAT satellite (van den Heuvel et al.,
1992; Rappaport et al., 1994; Kahabka and van den Heuvel, 1997).
Most of the known SSSs are located in the Large Magellanic Cloud,
Small Magellanic Cloud and M31. They typically emit
$10^{36}-10^{38}\,{\rm erg\,s^{-1}}$ in the form of very soft
X-rays, peaking in the energy range 20$-$100\,eV.

van den Heuvel et al. (1992) proposed a model that the relatively
massive WD sustains steady H-burning from a MS or subgiant donor star. They suggested that the
mass-accretion occurs at an appropriate rate, in the range of
$1.0-4.0\times10^{-7}\,M_{\rm \odot}{\rm yr}^{-1}$. Meanwhile, a WD
+ He star system has luminosity around $10^{37}-10^{38}\,{\rm
erg\,s^{-1}}$ when the He-burning is stable on the surface of the
WD, which is consistent with that of observed from SSSs.
Thus, WD + He star systems may also appear as SSSs before SN
explosions (Iben and Tutukov, 1994; Yoon and Langer, 2003; Wang et
al., 2009a). In addition, in the context of SSSs, the time that
elapses between the double WD merger and the SN explosion is about
$10^{5}$\,yr, and during this phase the merged object would look
like as a SSS (with $T\sim0.5-1\times10^{6}$\,K and $L_{\rm
X-ray}\sim10^{37}\,{\rm erg\,s^{-1}}$), which could provide a
potential test for the DD model (Yoon et al., 2007; Voss and Nelemans, 2008).
Note that the Galactic interstellar absorption and
circumstellar matter may play an important role in the obscuration
of X-rays.

Recently, Di Stefano (2010a,b) called attention to the fact that in
the galaxies of different morphological types there exists a
significant (up to 2 orders of magnitude) deficit of SSSs as
compared with expectations based on SN~Ia birthrates from the SD
model. Gilfanov and Bogd\'{a}n (2010) also obtained the same
conclusion, based on the study of the luminosity of elliptical
galaxies in the supersoft X-ray range. However, these authors did
not consider the binary evolution. A typical
binary in the SD model undergoes three evolutionary stages in order of time before
SN explosion, i.e. the wind phase, the supersoft X-ray source phase and
the recurrent nova phase, since the mass-accretion rate
decreases with time as the mass of the donor star decreases. The supersoft X-ray source phase
is only a short time (e.g. a few hundred thousand years), since the SD progenitor system spends a large part of lifetime in the wind
phase or recurrent nova phase on its way to SN explosion (e.g. Han and
Podsiadlowski, 2004; Meng et al., 2009; Wang et al., 2009a, 2010a; Hachisu et al., 2010; Meng and Yang, 2011a). Lipunov et al. (2011) also
considered that the theoretical SSS lifetimes and X-ray luminosities
have been overestimated.  

\subsection{Cataclysmic variables}
Cataclysmic variable stars (CVs) are stars that irregularly increase
in brightness by a large factor, then drop back down to a quiescent
phase (Warner, 1995). They consist of two component stars: a WD primary
and a mass donor star. CVs are usually divided into several
types, such as classical novae, recurrent novae, nova-like
variables, dwarf novae, magnetic CVs and AM CVns, etc (Warner, 1995). Among these subclasses of
CVs, recurrent novae and dwarf novae are the most probable candidates of SN Ia progenitors.

Recurrent novae have outbursts of about 4$-$9 magnitudes, and exhibit multiple outbursts at intervals of
10$-$80\,years (Warner, 1995). They
contain a massive WD and a relatively high mass-accretion rate (but below steady burning rate).
The evolution of the outburst is very fast. Since the heavy element
enhancement is not detected in recurrent novae, their WD mass is supposed
to increase after each outburst. Additionally, nova
outbursts require a relatively high
mass-accretion rate  onto a massive WD to
explain the recurring nova outbursts. Thus, these objects become some
of the most likely candidates of SN~Ia progenitors (Starrfield et al., 1985;
Hachisu and Kato, 2001). However, this class of objects are rare, with ten Galactic recurrent
novae, two in the Large Magellanic Cloud and a few in M31. Recurrent novae and SSSs differ in the
mass-accretion rate from a mass donor star onto the WD; SSSs have
steady nuclear burning on the surface of the WD, while recurrent
novae happen at rates that allow shell flashes.

By modeling the decline of the outburst light curves of some
recurrent novae (T CrB, RS Oph, V745 Sco and V3890 Sgr), Hachisu and
Kato (2001) suggested that these WDs are approaching the Ch mass and
will produce SNe~Ia. Recurrent nova systems like RS Oph have been
proposed as possible SN~Ia progenitors, based on the high mass of
the accreting WD. Patat et al. (2011a) investigated the circumstellar
environment of RS Oph and its structure, suggesting that the
recurrent eruptions might create complex structures within
the material lost by the donor star. This may establish a strong
link between RS Oph and the progenitor system of SN 2006X, for which
similar features have been detected.

Recurrent nova U Sco contains a WD of $M_{\rm WD}=1.55\pm0.24\,M_{\odot}$ and a secondary star with
$M_{\rm 2}=0.88\pm0.17\,M_{\odot}$ orbiting with a period $P_{\rm orb}\sim0.163$\,day (Thoroughgood et al., 2001).
The high mass of the WD implies that  U Sco is a strong
progenitor candidate of a SN~Ia (Thoroughgood et al., 2001; also see Hachisu et al., 2000). However, the
nebular spectra of U Sco displays that the relative abundance of
[Ne/O] is 1.69, which is higher than that of the typical [Ne/O] abundance
found in classical novae from CO WDs and suggests that U Sco has a
O-Ne-Mg WD (Mason, 2011). Thus, U Sco may not explode as a SN~Ia
but rather collapse to a neutron star by electron capture on $^{24}$Mg.

Dwarf novae have multiple outbursts ranging in brightness from 2 to 5 magnitudes,
and exhibit intervals from days to decades.
The lifetime of an outburst is typically from 2 to 20 days and is
related to the outburst interval. Dwarf nova outbursts are usually
attributed to the release of gravitational energy resulted from
an instability in the accretion disk or by sudden mass-transfer
via the disk (Warner, 1995). Observationally, there are a
number of dwarf novae in which the WD is about $1\,M_{\odot}$
(e.g. GK Per, SS Aur, HL CMa, U Gem, Z Cam, SY Cnc, OY Car,
TW Vir, AM Her, SS Cyg, RU Peg, GD 552 and IP Peg, etc). The
secondaries of these WD binaries are K or M stars ($<$$1\,M_{\odot}$).
A few of these systems
with early K type secondaries may have the WD mass close to the Ch
mass. It has been suggested that the mass-accretion rate onto
a WD during a dwarf nova outbursts can be sufficiently high to allow
steady nuclear burning of the accreted matter and growth of the WD
mass (King et al., 2003; Xu and Li, 2009; Wang et al., 2010a; Meng
and Yang, 2010a). However, whether dwarf nova outbursts can increase the mass of
a WD close to Ch mass is still a problem (e.g. Hachisu et al., 2010).

\subsection{Symbiotic systems}

Symbiotic systems are long-period binaries, consisting of
a RG and a hot object that is usually a WD (Truran and Cameron, 1971).
The hot object accretes and burns material from the RG star
via stellar wind in most cases, but could also be
RLOF in some cases. They usually show strong emission lines from surrounding circumstellar
material ionized by the hot component,
and low temperature absorption features from the RG.
Symbiotic systems are essential to understand the
evolution and interaction of detached and semi-detached binaries.
There are two distinct subclasses of symbiotic stars, i.e. the S-type
(stellar) with normal RG stars and orbital periods of about 1$-$15
years, and the D-type (dusty) with Mira primaries usually surrounded
by a warm dust shell and orbital periods longer than 10 years.
Symbiotic stars are thus interacting binaries with the longest
orbital periods. Tang et al. (2012) recently found a peculiar symbiotic system J0757 that
consists of an accreting WD and a RG. In quiescent phase, however, it doesn't show any signature of ``symbiotic''.
Thus, it is a missing population among symbiotic systems, which may contribute to a significant fraction of SN Ia.
Moreover, this object showed a 10 year flare in the 1940s, possibly from H-shell
burning on the surface of the WD and without significant
mass-loss. Therefore, the WD could grow effectively.

The presence of both the accreting WD and the RG
star makes symbiotic binaries a promising nursery for the production
of SNe~Ia. However, due to the low efficiency of matter
accumulation by a WD accreting material from the stellar wind, SN~Ia
birthrate from these symbiotic systems is relative low (e.g. Yungelson and Livio,
1998).

\subsection{Single low-mass He WDs}
The existence of a population of single low-mass He WDs (LMWDs;
$<$$0.45\,M_{\odot}$) is supported by some recent observations (e.g. Marsh et al., 1995; Kilic
et al., 2007). However, it is still unclear how to form single LMWDs.
It has been suggested that single LMWDs could be produced by
single old metal-rich stars that experience significant mass-loss
before the central He flash (Kalirai et al., 2007; Kilic et al., 2007).
However, the study of the initial-final mass
relation for stars by Han et al. (1994) implied that only LMWDs with
masses larger than $0.4\,M_{\odot}$ might be produced from such a
single star scenario, even at high metallicity environment (Meng et
al., 2008). Thus, it would be difficult to conclude that single
stars can produce LMWDs of $\sim$$0.2\,M_{\odot}$.

Justham et al. (2009) inferred an attractive formation scenario for
single LMWDs, which could be formed in binaries where their
companions have exploded as SNe~Ia. Wang and Han (2010d) recently
found that the surviving companions of the old SNe~Ia from the WD +
MS and WD + RG channels have low masses, providing a possible
way to explain the formation of the population of single LMWDs (see
also Meng and yang, 2010c). Conversely, the observed single LMWDs may
provide evidence that at least some SN~Ia explosions have occurred
with non-degenerate donors (such as MS or RG donors). We note that Nelemans and Tauris (1998)
also proposed an alternative scenario to form single LMWDs from a
solar-like star accompanied by a massive planet, or a brown dwarf,
in a relatively close binary orbit.

\subsection{Hypervelocity stars}
In recent years, hypervelocity stars (HVSs) have been observed in
the halo of the Galaxy. HVSs are stars with velocities so high that
they are able to escape the gravitational pull of the Galaxy.
However, it is still not clear how to form HVSs (for a review see
Tutukov and Fedorova, 2009). It has been suggested that such HVSs
can be formed by the tidal disruption of a binary through
interaction with the super-massive black hole (SMBH) at the Galactic
center (GC) (Hills, 1988; Yu and Tremaine, 2003; Zhang et al.,
2010).

The first three HVSs have only recently been discovered
serendipitously (e.g. Brown et al., 2005; Hirsch et al., 2005;
Edelmann et al., 2005). Up to now, about 17 confirmed  HVSs have been
discovered in the Galaxy (Brown et al., 2009; Tillich et al., 2009),
most of which are B-type stars, probably with masses ranging from 3
to 5\,$M_\odot$ (Brown et al., 2005, 2009; Edelmann et al., 2005).
The HVS B-type stars are demonstrated
short-lived B-type stars at 50$-$100\,kpc distances that are significantly unbound
based on radial velocity alone.  Their observed properties (ages, flight
times, latitude distribution) are consistent with the Galactic center
ejection scenario (Brown et al., 2012b).
One HVS, HE 0437-5439, is known to be an apparently normal early
B-type star. Edelmann et al. (2005) suggested that the star could
have originated in the Large Magellanic Cloud, since it is much
closer to this galaxy ($\sim$18\,kpc) than to the GC (see also
Przybilla et al., 2008). Li et al. (2012) recently
reported 13 metal-poor F-type HVS candidates which are selected from 370,000
stars of the data release 7 of the Sloan Digital Sky Survey. With a
detailed analysis of the kinematics of these stars, they claimed that
seven of them were likely ejected from the GC or the Galactic disk,
four neither originated from the GC nor the Galactic disk, and the
other two were possibly ejected from either the Galactic disk or
other regions.

At present, only one HVS, US 708, is an extremely He-rich sdO star
in the Galactic halo, with a heliocentric radial velocity of
+$708\pm15$\,km/s. Hirsch et al. (2005) speculated that US
708 was formed by the merger of two He WDs in a close binary induced
by the interaction with the SMBH in the GC and then escaped.
Recently, Perets (2009) suggested that US 708 may have been
ejected as a binary from a triple disruption by the SMBH, which
later on evolved and merged to form a sdO star. However, the evolutionary lifetime of
US 708 is not enough if it originated from the GC.
Wang and Han (2009)
found that the surviving companions from the He star donor channel
have a high spatial velocity ($>$400\,km/s) after a SN explosion,
which could be an alternative origin for HVSs, especially for HVSs
such as US 708 (see also Justham et al., 2009). Considering the
local velocity nearby the Sun ($\sim$220\,km/s), Wang and Han (2009)
found that about 30$\%$ of the surviving companions may be observed
to have velocity above 700\,km/s. In addition, a SN asymmetric explosion may
also enhance the velocity of the surviving companion. Thus, a
surviving companion star in the He star donor channel may have a
high velocity like US 708.

\section{Origin of SN~Ia diversity}

SNe~Ia have been successfully used as cosmological distance
candles, but there exists spectroscopic diversity among SNe~Ia that is
presently not well understood, nor how this diversity is linked to the
properties of their progenitors (e.g. Branch et al., 1995; Livio, 2000).
When SNe~Ia are applied as distance indicators, the Phillips relation
is adopted (i.e. the luminosity-width relation; brighter SNe Ia
have wider light curves), which implies that SN~Ia luminosity is mainly determined by one
parameter. In an attempt to quantify the rate of spectroscopically
peculiar SNe~Ia in the existing observed sample, Branch et al.
(1993) compiled a set of 84 SNe~Ia and found that about $83\% -
89\%$ of the sample are normal. According to the study of Li et
al. (2001), however, only $64\%\pm 12\%$ of the observed SNe~Ia are
normal in a volume-limited search.\footnote{There is increasing evidence showing
that even the normal SNe Ia exhibit diversity in their spectral features
(e.g. Branch et al., 2009; Wang et al., 2009c; Blondin et al., 2012).
Wang et al. (2009c) investigated
158 relatively normal SNe Ia by dividing them into two
groups in terms of the expansion velocity inferred from the absorption minimum of the Si${\rm II}$ ${\rm \lambda}6355$ line around maximum light. They claimed that, one group ``Normal'' consists of  SNe Ia with an average expansion velocity $10,600\pm400\rm km/s$, but another group ``HV'' consists of objects with higher velocities $\sim$$11,800\rm km/s$.
The HV SNe~Ia are found
to prefer a smaller extinction ratio $R_{V}$ (relative to the Normal ones), which might suggest the presence of circumstellar material (see Sect. 3.7).} The total rate of peculiar SNe~Ia
could be as high as $36\%\pm 9\%$; the rates are $16\%\pm 7\%$ and
$20\%\pm 7\%$ for SN 1991bg-like objects and SN 1991T-like objects,
respectively. SN 1991bg-like objects both rise to
their maximum and decline more quickly, and are sub-luminous
relative to normal SNe Ia, whereas SN 1991T-like objects both
rise to their maximum and decline more slowly, and are more luminous
relative to normal SNe Ia. These two types of
peculiar events obey the luminosity-width relation.
However, a subset of SNe~Ia apparently
deviate from the luminosity-width relation, e.g.
some were observed with exceptionally high luminosity or extremely low luminosity, which may have progenitors with
masses exceeding or below the standard Ch mass limit (e.g.
Howell et al., 2006; Foley et al., 2009). This implies that at least some SNe~Ia can be produced
by a variety of different progenitor systems, and probably suggests that SN~Ia luminosity is not
the single parameter of the light curve shape.

It has been suggested that the amount of $^{\rm 56}$Ni formed during a
SN Ia explosion dominates its maximum luminosity (Arnett,
1982), but the origin of the variation of the amount of $^{\rm
56}$Ni for different SNe~Ia is still unclear (the derived
$^{\rm 56}$Ni masses for different SNe Ia could vary by a factor of ten; Wang et al., 2008a). Many efforts have been
paid to solve this problem. Umeda et al. (1999) suggested that the
average ratio of carbon to oxygen (C/O) of a WD at the moment of a SN
explosion is the dominant parameter for the Phillips relation, i.e. the
higher the C/O ratio, the larger the amount of $^{56}$Ni, and then
the higher the maximum luminosity (see also Meng and Yang, 2011b). However, 3D simulations by
R\"{o}pke and Hillebrandt (2004) suggest that different C/O ratios
have a negligible effect on the amount of $^{56}$Ni produced.
At present, the studies from the explosion models of SNe Ia indicate that the number of
ignition points at the center of WDs or the transition density from deflagration to detonation
dominates the production of $^{56}$Ni, and consequently the maximum luminosity (e.g. Hillebrandt and Niemeyer, 2000; H\"{o}flich et
al.,  2010; Kasen et al., 2010).

It was claimed that the ignition
intensity (the number of ignition points) in the center of WDs is a
useful parameter in interpreting the Phillips relation (Hillebrandt
and Niemeyer, 2000). Based on the SD model, Lesaffre et al. (2006) carried out a systematic
study of the sensitivity of carbon ignition conditions for the Ch mass
WDs on various properties, and
claimed that the central density of a WD at the carbon ignition may be the
origin of the scatter of the maximum luminosity.
This suggestion was further supported by detailed multi-dimensional numerical
simulations of SN explosions (Krueger et al., 2010). We note that the WD cooling time
before mass-accretion is less than 1\,Gyr in the simulations of Lesaffre et al. (2006) and
Krueger et al. (2010). However, there are SNe Ia with the delay times
$\sim$10 Gyr in observations. The WDs with such a long cooling
time may become more degenerate before the onset of the mass-accretion
phase. Some other processes, such as carbon and oxygen separation
or crystallization, may occur and dominate the properties of the
CO WD (Fontaine et al., 2001). How the extremely degenerate
conditions affect the properties of SNe Ia still remains unclear. The
suggestion of Lesaffre et al. (2006) should be checked carefully
under extremely degenerate conditions.
Adopting the WD mass-accretion process in Lesaffre et al. (2006),
Chen et al. (2012) recently studied the evolution of various CO WDs
from the onset of mass-accretion to carbon ignition at Ch mass limit. The
study shows that the carbon ignition generally occurs at the center for hot low-mass
CO WDs but off-center for cool massive ones, which may
provide more information for the explosion models of SNe Ia.

Some numerical and synthetical results showed that the
metallicity may have an effect on the final amount of $^{56}$Ni, and
thus the maximum luminosity of SNe~Ia (Timmes et al., 2003;
Podsiadlowski et al., 2006; Bravo et al., 2010). There is also some
other evidence of the correlation between the properties of SNe~Ia
and metallicity from observations (e.g. Branch and Bergh, 1993;
Hamuy et al., 1996; Wang et al, 1997; Gallagher et al., 2008; Sullivan, 2006; Howell et al., 2009a;
Sullivan et al., 2010). Podsiadlowski et al. (2006) introduced
metallicity as a second parameter that affects the light curve shape.
For a reasonable range of metallicity, this may account for the
observed spread in the Phillips relation. Since metallicity in
the Universe has evolved with time, this introduces an undesirable
evolutionary effect in the SN~Ia distance method, which could mimic
the effect of an accelerating Universe. We also note that
Maeda et al. (2010) argued that the origin of spectral
evolution diversity in  SNe Ia can be understood by an asymmetry
in the SN explosion combined with the observer's
viewing angle. Moreover, Parrent et
al. (2011) investigated the presence of C${\rm II}$ ${\rm
\lambda}6580$ in the optical spectra of 19 SNe~Ia. Most of the objects in
their sample that exhibit C${\rm II}$ ${\rm \lambda}6580$ absorption
features are of the low-velocity gradient subtype. This study
indicates that the morphology of carbon-rich regions is consistent
with either a spherical distribution or a hemispheric asymmetry,
supporting the idea that SN~Ia diversity may be a result of
off-center ignition coupled with observer's viewing angle.

\section{Impacts of SN~Ia progenitors on some fields}
The identification of SN~Ia progenitors also has important impacts
on some other astrophysical fields, e.g. cosmology, the evolution of
galaxies, SN explosion models and binary
evolution theories, etc (e.g. Branch et al., 1995; Livio, 2000).

\textbf{Cosmology.} It is feasible to improve SNe~Ia as mature
cosmological probes, since the dominant systematic errors are clear,
which include photometric calibration, selection effects, reddening
and population-dependent differences, etc. In the next decade, SNe~Ia are
proposed to be cosmological probes for testing the evolution of the
dark energy equation of state with time (Howell et al., 2009b). The
use of SNe~Ia as one of the main ways to determine the Hubble
constant ($H_0$) and cosmological parameters (e.g.
\textbf{$\Omega_{M}$} and \textbf{$\Omega_{\Lambda}$}; Riess et al., 1998; Perlmutter et al.,
1999),
requires our understanding of the evolution of the luminosities and
birthrates of SNe~Ia with cosmic epoch. Both of these depend on the
nature of their progenitors. Meanwhile, the evolution of the
progenitor systems or a changing mix of different progenitors may
bias cosmological inferences. For a recent review of this field see Howell (2011).

\textbf{Galaxy evolution.} Aside from cosmology, the
evolution of galaxies depends on the radiative, kinetic energy,
nucleosynthetic outputs (e.g. Kauffmann et al., 1993; Liu et al., 2012a)
and the birthrates of SNe~Ia with time, which all
depend on the nature of the progenitor systems. SNe Ia are also
laboratories for some extreme physics, e.g. they are accelerators
of cosmic rays and as sources of kinetic energy
in galaxy evolution processes (e.g. Helder et al., 2009; Powell
et al., 2011).  Especially, SNe~Ia
regulate galactic and cluster's chemical evolution. Due to the main
contribution of iron to their host galaxies, SNe Ia are a key part of our
understanding of galactic chemical evolution (e.g.
Greggio and Renzini, 1983; Matteucci and Greggio, 1986).
The existence of young and old
populations of SNe~Ia suggested by recent observations may have an important
effect on models of galactic chemical evolution, since they would
return large amounts of iron to the interstellar medium either much earlier
or much later than previously thought.

\textbf{Explosion models.} SNe Ia provide natural laboratories for studying the
physics of hydrodynamic and nuclear processes with extreme
conditions. The link between the progenitor models
and the explosion models is presently one of the weakest points in
our understanding of SNe~Ia (Hillebrandt and Niemeyer, 2000).
Due to some uncertainties that still exist in the SN explosion
mechanism itself, a knowledge of the initial conditions and the
distribution of matter in the environment of the exploding star is
essential for our understanding of SN explosion, e.g. the ignition
density may depend on the initial WD mass, the age of the
progenitor, the metallicity and the treatment of rotation in the
progenitor. Moreover, different progenitor models may lead to
different WD structures before SN explosion. Lu et al. (2011)
recently studied the properties of the Tycho's SNR.
They estimated the parameters of the binary system before the SN explosion,
which may shed lights on the possible explosion models.

\textbf{Binary evolution theories.} The identification of
SN~Ia progenitors, coupled with observationally determined SN~Ia
birthrates and delay times will help to place meaningful constraints
on some theories of binary evolution, e.g. the mass-transfer between
two stars, the mass-accretion efficiency of WDs, etc (e.g. Hachisu et al., 
1996; Han and Podsiadlowski, 2004; Wang et al., 2009a). Especially, it
is possible that the CE efficiency parameter may be constrained (e.g. Meng et al, 2011),
which is important in binary evolution and BPS studies.

\section{Summary}
In this article, various progenitor models proposed in the
literatures are reviewed, including some variants of SD and DD models.
We addressed some observational ways to test the current
progenitor models and introduced some observed objects that may be related to  the
progenitors and the surviving companion stars of SNe Ia.  We also discussed the impacts of SN~Ia
progenitors on some fields. The origin of the observed SN~Ia diversity is still unclear.
It seems likely that SNe~Ia can be produced by a variety of different progenitor systems, perhaps explaining
part of the observed diversity. SN asymmetric explosion coupled with observer's viewing angle may also produce the diversity.
Additionally, the metallicity of progenitors may
be a second parameter that affects the light curve shape of SNe Ia.

At present, the SD model is the most widely accepted SN Ia progenitor model.
The advantages of this model can be summarized as follows:
\begin{enumerate}[(1)]
\item The SD model is in excellent agreement with the observed light curves and spectroscopy of SNe Ia, and this model may explain the similarities of most SNe Ia.
\item Observationally, there is increasing evidence indicating that some SNe~Ia may come from the SD model (e.g. the signatures of gas outflows from some SN~Ia progenitor systems, the wind-blown cavity in SN remnant, and the early optical and UV emission of SNe Ia, etc). In addition, the SD model may be compatible with some recent observations (e.g. the lack of H or He seen in nebular spectra of SNe Ia, and the upper limits from SN Ia radio and X-ray detection, etc) by considering the spin-down time.
\item There are some SD progenitor candidates in observations, e.g. supersoft X-ray sources, recurrent novae, dwarf novae and symbiotic systems, etc. Meanwhile, a number of high mass WDs that have been accreting from a non-degenerate companion star have been found.
\item The observed single low-mass He WDs and hypervelocity He stars may be explained by the surviving companion stars predicted in the SD model.
\item SNe Ia with long delay times can be understood by the WD + MS and WD + RG channels. In contrast, SNe Ia with short delay times may consist of systems with a He donor star in the WD + He channel, or even a massive MS donor star in the WD + MS channel.
\item Besides the DD model, these observed super-luminous SNe~Ia can also  be produced by the SD model by considering the effects of rapid differential rotation on the accreting WD.
\end{enumerate}

However, the SD model is still suffering some problems from both theoretically and
observationally that need to be resolved:
\begin{enumerate}[(1)]
\item The optically thick wind assumption, widely adopted in the studies of the SD model, is in doubt for very low metallicity; the low-metallicity threshold for SNe Ia predicted by theories has not been found in observations.
\item It is still difficult to reproduce the observed birth rates and delay times of SNe~Ia. This suggests that we need a better understanding of mass-accretion onto WDs.
\item There is still no conclusive proof that any individual object is the surviving companion star of a SN~Ia, which is predicted by the SD model. A likely surviving companion star for the progenitor of Tycho's SN has been identified, but the claim is still controversial.
\end{enumerate}

Although a DD merger is thought to experience an accretion-induced
collapse rather than a thermonuclear explosion,
any definitive conclusion about the DD model is currently premature:
\begin{enumerate}[(1)]
\item There are some parameter ranges in which the accretion-induced collapse can be avoided. Recent simulations indicate that the violent mergers of two massive WDs can closely resemble normal SN Ia explosion with the assumption of the detonation formation as an artificial parameter, although these mergers may only contribute a small fraction to the observed population of normal SNe Ia.
\item This model can naturally reproduce the observed birthrates and delay times of SNe Ia and may explain the formation of some observed super-luminous SNe~Ia.
\item This model can explain the lack of H or He seen in the nebular spectra of SNe Ia.
\item Recent observational studies of SN 2011fe seem to favor a DD progenitor. In addition, there is no signal of a surviving companion star from the central region of SNR 0509-67.5 (the site of a SN Ia explosion whose light swept Earth about 400 years ago), which may indicate that the progenitor for this particular SN Ia is a DD system.
\item Some observed double WD systems may have the total mass larger than the Ch mass, and possibly merge within the Hubble-time, although there are not enough double WD systems to reproduce the observed SN Ia birthrates in the context of the DD model.
\end{enumerate}

Some variants of the SD and DD models have been
proposed to explain the observed diversity of SNe~Ia:
\begin{enumerate}[(1)]
\item The sub-luminous 1991bg-like objects may be explained by the sub-Ch mass model.
\item The unusual properties of 2002ic-like objects can be understood by the delayed dynamical instability model.
\item The spin-up/spin-down model may provide a route to explain the similarities and the diversity observed in SNe~Ia.
\item The core-degenerate model could form a massive WD with super-Ch mass that might explode as a super-luminous SN~Ia.
\item The collisions between two WDs in dense environments could also potentially lead to sub-luminous SN Ia explosions.
\item The mechanism of WDs exploding near black holes is also a potential progenitor model for thermonuclear runaway, despite of the expected low rate when a WD passes near a black hole.
\end{enumerate}

To set further constraints on SN Ia progenitor models, large samples
of SNe~Ia with well-observed light curves and spectroscopy in nearby
galaxies are required to establish the connection of SN~Ia
properties with the stellar environments of their host galaxies.
Many new surveys from ground and space have been proposed to
make strides in SN~Ia studies , e.g. Palomar Transient Factory, Skymapper, La Silla
QUEST, Pan-STARRS, the Dark Energy Survey, Large Synoptic Survey
Telescope, the Joint Dark Energy Mission and the Gaia Astrometric Mission, etc (Howell
et al., 2009; Altavilla et al., 2012). These surveys will
allow comparisons via large SN Ia subsamples, and start to
connect SN~Ia progenitors with the observed features of SN explosions
themselves, and thus to unveil the nature of SN Ia progenitors.

\section*{Acknowledgments}
We acknowledge useful comments and suggestions from Shuangnan Zhang and Stephen Justham.
We also thank Simon Jeffery, Xiaofeng Wang, Xiangcun Meng, Xuefei Chen and Zhengwei Liu for their helpful discussions.
This work is supported by the National Natural Science Foundation of
China (Grant Nos. 11033008 and 11103072), the National
Basic Research Program of China (Grant No. 2009CB824800), the
Chinese Academy of Sciences (Grant No. KJCX2-YW-T24), the
Western Light Youth Project and Youth Innovation Promotion
Association of the Chinese Academy of Sciences.

\end{document}